\documentclass[12pt,preprint]{aastex}  

\usepackage{graphicx}

\begin{document}

\title{Population III star formation in a $\Lambda$WDM universe}
\author{Brian W. O'Shea\altaffilmark{1} \& Michael L. Norman\altaffilmark{2}}

\altaffiltext{1}{Theoretical Astrophysics (T-6), Los Alamos National
Laboratory, Los Alamos, NM 87545; bwoshea@lanl.gov}

\altaffiltext{2}{Center for Astrophysics and Space Sciences,
University of California at San Diego, La Jolla, CA 
92093; mnorman@cosmos.ucsd.edu}

\begin{abstract}
In this paper we examine aspects of primordial star formation in a gravitino
warm dark matter universe with a cosmological constant ($\Lambda$WDM).  We
compare a set of simulations using a single cosmological realization
but with a wide range of warm dark matter particle masses which have not yet
been conclusively ruled out by observations.  
The addition of a warm dark matter component to the initial power spectrum
results in a delay in the collapse of high density gas at the center of the most
massive halo in the simulation and, as a result, an increase in the virial mass of 
this halo at the onset of baryon collapse.  Both of these effects become more 
pronounced as the warm dark matter
particle mass becomes smaller.  A cosmology using a gravitino warm dark matter power
spectrum assuming a particle mass of m$_{WDM} \simeq 40$~keV is effectively
indistinguishable from the cold dark matter case, whereas the m$_{WDM} \simeq 15$~keV
case delays star formation in the parent halo by $\simeq 10^8$~years.   
There is remarkably little scatter between simulations in 
the final properties of
the primordial protostar which forms at the center of the halo, possibly due to the
overall low rate of halo mergers which is a result of the WDM power spectrum.  The
detailed evolution of the collapsing halo core in two representative WDM cosmologies
is described.  At low densities (n$_{b} \la 10^5$~cm$^{-3}$), the evolution of the 
two calculations is qualitatively similar, but occurs on significantly different 
timescales, with the halo in the lower particle mass calculation taking much
longer to evolve over the same density range and reach runaway collapse. 
Once the gas in the center of the halo reaches relatively high
densities (n$_{b} \ga 10^5$~cm$^{-3}$) the overall evolution is essentially 
identical in the two calculations.
\end{abstract}


\maketitle

\section{Introduction}\label{Intro}

The current canonical view of cosmological structure formation, which assumes 
that dark matter halos and their contents form in a bottom-up (hierarchical) 
fashion in a universe with cold dark matter and a cosmological constant 
($\Lambda$CDM), has been successful in predicting the formation and evolution 
of  large scale structure, as shown in recent years by extensive surveys of the
local universe, such as the 2dF and Sloan Digital Sky 
Survey~\markcite{bahcall99,2001MNRAS.327.1297P, 2004ApJ...606..702T}({Bahcall} {et~al.} 1999; {Percival} {et~al.} 2001; {Tegmark} {et~al.} 2004).  It is 
acknowledged, however, that there appear to be flaws in the $\Lambda$CDM
scenario.  
Observations of the central regions of galaxies show that the central dark matter 
density of these halos is significantly lower than that predicted by 
theory~\markcite{1997ApJ...490..493N, 2004MNRAS.349.1039N, 1999MNRAS.310.1147Mb,2000ApJ...544..616G,
2000ApJ...531L.107S,2000AJ....119.1579V,2001ApJ...561...35D}({Navarro}, {Frenk}, \&  {White} 1997; {Navarro} {et~al.} 2004; {Moore} {et~al.} 1999b; {Ghigna} {et~al.} 2000; {Swaters}, {Madore}, \&  {Trewhella} 2000; {van den Bosch} {et~al.} 2000; {Dalcanton} \& {Hogan} 2001).
  In addition, the
CDM model predicts the formation of a larger number of dwarf galaxies in the local 
group than is observed (by an order of magnitude), and also suggests
that many dwarf galaxies will form in the cosmic voids -- a prediction that has 
not been verified observationally
~\markcite{1999ApJ...524L..19M,2004MNRAS.353..639W,1999ApJ...522...82K,2001ApJ...556...93B,
2001ApJ...557..495P}({Moore} {et~al.} 1999a; {Willman} {et~al.} 2004; {Klypin} {et~al.} 1999; {Bode}, {Ostriker}, \&  {Turok} 2001; {Peebles} 2001).  In addition there is the ``angular momentum'' problem of CDM halos, where gas cools
at early times into small mass halos, leading to massive, low angular momentum 
cores in galaxies that are not observed~\markcite{2001ApJ...551..608S}({Sommer-Larsen} \& {Dolgov} 2001).  The 
formation of disk-dominated or disk-only galaxies may be impeded by bulge formation 
in the CDM model due to high merger 
rates~\markcite{2004ApJ...607..688G,2005RMxAC..23..101K}({Governato} {et~al.} 2004; {Kormendy} \& {Fisher} 2005).  Finally, there have been 
observations of significant numbers of dwarf galaxies forming after the 
larger Lyman break galaxies, which is not what one would expect in a hierarchical 
clustering scenario~\markcite{2001MNRAS.323..795M}({Metcalfe} {et~al.} 2001).  However, at large scales 
($\ga 1$~Mpc) the CDM model 
seems to describe the evolution of the universe and the structure
within it quite well. 

It is unclear whether the disagreement between the CDM theory of structure 
formation and current observations is due to an incorrect comparison of 
simulation and observational data or an absence of essential baryonic 
processes in the simulations.  \markcite{2004ApJ...609..482K}{Kravtsov}, {Gnedin}, \&  {Klypin} (2004) suggest that the 
observed disagreement may be due to incorrectly interpreting dark matter-only 
simulations, and introduce a model that appears to correctly reproduce
the abundance and other properties of observed galactic satellites.  
\markcite{2005ApJ...622L..21M}{Minchin} {et~al.} (2005) describe the discovery of an optically dark object 
in the Virgo cluster found via deep HI survey.  This object has a
neutral hydrogen gas mass of $\sim 10^8$~M$_\odot$ and a velocity width of 
$\Delta V_{20} = 220$~km/s.  Measurements of the spatial extent of the object 
suggest a minimum dynamical mass of $\sim 10^{11}$~M$_\odot$.  The Tully-Fisher 
relation suggests that a galaxy with this velocity width should be 12th magnitude
or brighter in optical bands; however, deep imaging has failed to find an 
optical counterpart down to a surface brightness of 27.5 B mag/arcsec$^2$.  The 
discovery of this object implies the suppression of star formation in 
galaxy-sized halos, and the authors suggest that a population of these ``dark'' 
objects may be found in further deep, high-resolution HI surveys.  Finally, 
\markcite{2002ApJ...572...25D}{Dalal} \& {Kochanek} (2002) and~\markcite{2002ApJ...565...17C}{Chiba} (2002) argue that 
abundances of inferred dark matter substructure in several galactic lens 
systems is consistent with predictions in standard CDM models.  It is clear that
there are aspects of the $\Lambda$CDM paradigm that have yet to be fully 
explored, and disagreements between theory and observation to be reconciled,
before cold dark matter can be conclusively ruled out.

If one assumes that there is in fact a flaw with the cold dark matter scenario, 
the problem is then to devise some physical explanation for the apparent 
lack of power on small scales while also retaining the desirable qualities of 
the CDM model on large scales.  Many models have been proposed that can provide 
this suppression.  In this paper we discuss the ramifications of a general class
of these models, referred to as ``warm dark matter,'' on the formation of 
Population III stars.  These models predict exponential damping of the linear 
power spectrum on small length scales \markcite{1986ApJ...304...15B,1996ApJ...458....1C}({Bardeen} {et~al.} 1986; {Colombi}, {Dodelson}, \&  {Widrow} 1996).
 The effects of a gravitino warm dark matter cosmology are discussed 
by~\markcite{2001ApJ...556...93B}{Bode} {et~al.} (2001), who derive the relevant linear perturbation theory 
and perform several N-body calculations of warm dark matter cosmologies to understand the
general effects of suppression of power on small scales.  They find that replacing 
cold dark matter with warm dark matter results in the smoothing of massive halo cores,
which lowers core densities and increases core radii, lowers the characteristic density
of low-mass halos, reduces the overall total number of low-mass halos, suppresses the
number of low-mass satellite halos in high-mass halos, and results in the formation
of low-mass halos almost entirely within caustic sheets or filaments connecting
larger halos -- voids are almost completely empty, in contrast to CDM.  They also find
that low-mass halos tend to form at late times, in a top-down process (as opposed to 
the bottom-up process of halo formation one would expect from a CDM cosmology),
and that halo formation is suppressed overall at early times (high redshifts), with an 
increased evolution
of halos at low redshifts relative to the CDM model (this result was later confirmed by
\markcite{2003MNRAS.345.1285K}{Knebe} {et~al.} (2003), though see \markcite{2003Ap&SS.284..341G}{G{\"o}tz} \& {Sommer-Larsen} (2003), who suggest that
some of the observed effects are due to inappropriate initial conditions).  
Furthermore, they argue that  constraints based on the
observed halo mass function and its evolution suggest that a reasonable minimum warm 
dark matter particle mass would be 1~keV.  

Constraints on the minimum mass of a warm dark matter particle have been placed by 
various other groups as well.  \markcite{2000ApJ...543L.103N}{Narayanan} {et~al.} (2000) use the clustering
properties of Lyman-$\alpha$ forest absorbers to suggest a lower limit of 
m$_{WDM} = 0.75$~keV.  \markcite{2001ApJ...558..482B}{Barkana}, {Haiman}, \&  {Ostriker} (2001) use
an extended Press-Schechter model to constrain warm dark matter based on 
observations of cosmological reionization.  They calculate that in order
for super massive black holes to exist at $z \simeq 6$, and if massive
galaxies are responsible for the nearly complete reionization of the universe
by the same redshift, a reasonable minimum mass for a warm dark matter 
particle is m$_{WDM} \geq 1.2$~keV.  \markcite{2003ApJ...593..616S}{Somerville}, {Bullock}, \&  {Livio} (2003) observe that
the WMAP Year 1 polarization result~\markcite{kogut03}({Kogut} {et~al.} 2003) requires early structure formation
and may place a more stringent constraint on m$_{WDM}$~($\gg 1~$keV).  The lowering of
the optical depth with the WMAP Year 3 data release \markcite{2006astro.ph..3450P}({Page} {et~al.} 2006) 
weakens the strength of this argument.
\markcite{2005PhRvD..71f3534V}{Viel} {et~al.} (2005) use a combination of WMAP Year 1 and Lyman-$\alpha$ forest 
power spectrum data to constrain m$_{WDM}$, finding that m$_{WDM} \ga 550$~eV 
for early decoupled thermal relics and m$_{WDM} \ga 2$~keV for sterile neutrinos.  
Finally, \markcite{abazajian05}{Abazajian} (2005a) uses a combination of CMB power spectra (from 
multiple experiments) with 3D galaxy power spectra and 1D Lyman-$\alpha$ forest 
power spectra from the Sloan Digital Sky Survey to obtain a constraint on the WDM 
particle mass of $1.7$~keV~$<$~m$_s$~$< 8.2$~keV, assuming that the warm dark 
matter particle is a sterile neutrino which
suppresses power on small scales in a somewhat different way than the power spectrum
described by \markcite{2001ApJ...556...93B}{Bode} {et~al.} (2001).  The upper limit in this work comes from
constraints given by sterile neutrino decay rates \markcite{2001ApJ...562..593A}({Abazajian}, {Fuller}, \&  {Tucker} 2001),
and can be disregarded in the gravitino case.  The lower limit stated by 
Abazajian for gravitino dark matter is $0.5$~keV (95\% CL).

The effects of a warm dark matter particle on the formation of the first 
generation of stars in the universe may be profound.  The WDM particle masses 
which have not yet been ruled out by observation correspond to mass and spatial 
scales comparable to those of the halos in which Population III stars form 
(M$_{h} \sim 10^6$~M$_\odot$, $R_h \sim 100$ pc (proper) at $z \sim 20$).  WMAP 
Year 3 observations of CMB polarization implies a significant electron optical depth 
($\tau_e = 0.09 \pm 0.03$), which suggests that reionization could have begun
at redshifts comparable to the epoch of Pop III star formation in a $\Lambda$CDM 
universe \markcite{2006astro.ph..3450P,ABN02,2002ApJ...564...23B}({Page} {et~al.} 2006; {Abel}, {Bryan}, \& {Norman} 2002; {Bromm}, {Coppi}, \&  {Larson} 2002).  The effect of a warm dark 
matter particle would generally be to delay the formation of cosmological 
structure.  This has been examined in a statistical way by using techniques such 
as the extended Press Schechter (EPS) formalism~\markcite{2003ApJ...593..616S}({Somerville} {et~al.} 2003), but 
the details can only be studied accurately with the use of high-resolution 
numerical simulations.

To this end, \markcite{2003ApJ...591L...1Y}{Yoshida} {et~al.} (2003b) perform N-body+SPH cosmological 
simulations of structure formation in the early universe of a $\Lambda$CDM 
model and a $\Lambda$WDM gravitino model with a warm dark matter particle mass of 10 keV
assuming the \markcite{2001ApJ...556...93B}{Bode} {et~al.} (2001) transfer function.  
They find that, as expected, the power spectrum cutoff results in an absence of 
low-mass halos, which makes the formation of Population III stars very 
inefficient at redshifts comparable to the epoch of reionization implied by 
the WMAP Year 1 polarization result.  They suggest that, based on this polarization
observation, any successful warm dark 
matter model will have a particle mass greater than 10 keV.  Their conclusions, 
while valid, do not sample the parameter space of allowed warm dark matter 
particle masses, and the high dark matter and gas particle masses in
their simulations (m$_{DM} \simeq 1000$~M$_\odot$, m$_{b} \simeq 160$~M$_\odot$) 
do not allow direct examination of the Population III star-forming mass scales.
Additionally, the WMAP Year 3 polarization result reduces the strength of their
argument concerning limits on the WDM particle mass.

In this paper we address questions raised by Yoshida et al. in a higher level of 
detail, by both simulating a range of plausible warm dark matter particle masses 
that have not yet been ruled out by observation and by resolving spatial and mass
scales that allow direct examination of the formation of Population III protostellar 
cloud cores.  In Section~\ref{methodology} we discuss Enzo, the adaptive mesh 
refinement cosmological code used to carry out the simulations 
(Section~\ref{enzocode}), some general effects of the Bode et al. warm 
dark matter cosmological model (Section~\ref{wdmmodel}), and finally the setup 
of the series of cosmological simulations (Section~\ref{simsetup}).  In 
Section~\ref{results} we present our results, first showing some general effects of 
the variation of the warm dark matter particle mass (Section~\ref{results-general}),
comparing the detailed evolution of the halo cores within two 
representative warm dark matter models (Section~\ref{results-rep}), and
examining the top-down formation of a halo which eventually forms a Population III 
protostar in a universe with a relatively low WDM particle mass 
(Section~\ref{results-topdown}).  Finally, we 
discuss this work in Section~\ref{discuss} and provide a short summary of results 
in Section~\ref{summary}.

\section{Methodology}\label{methodology}

\subsection{The Enzo code}\label{enzocode}

`Enzo'\footnote{http://cosmos.ucsd.edu/enzo/} is a publicly available, extensively tested 
adaptive mesh refinement
cosmology code developed by Greg Bryan and others \markcite{bryan97,bryan99,norman99,oshea04,
2005ApJS..160....1O}({Bryan} \& {Norman} 1997a, 1997b; {Norman} \& {Bryan} 1999; {O'Shea} {et~al.} 2004, 2005).
The specifics of the Enzo code are described in detail in these papers (and references therein),
but we present a brief description here for clarity.

The Enzo code couples an N-body particle-mesh (PM) solver \markcite{Efstathiou85, Hockney88}({Efstathiou} {et~al.} 1985; {Hockney} \& {Eastwood} 1988) 
used to follow the evolution of a collisionless dark
matter component with an Eulerian AMR method for ideal gas dynamics by \markcite{Berger89}{Berger} \& {Colella} (1989), 
which allows high dynamic range in gravitational physics and hydrodynamics in an 
expanding universe.  This AMR method (referred to as \textit{structured} AMR) utilizes
an adaptive hierarchy of grid patches at varying levels of resolution.  Each
rectangular grid patch (referred to as a ``grid'') covers some region of space in its
\textit{parent grid} which requires higher resolution, and can itself become the 
parent grid to an even more highly resolved \textit{child grid}.  Enzo's implementation
of structured AMR places no fundamental restrictions on the number of grids at a 
given level of refinement, or on the number of levels of refinement.  However, owing 
to limited computational resources it is practical to institute a maximum level of 
refinement, $\ell_{max}$.  Additionally, the Enzo AMR implementation allows arbitrary 
integer ratios of parent
and child grid resolution, though in general for cosmological simulations (including the 
work described in this paper) a refinement ratio of 2 is used.

Since the addition of more highly refined grids is adaptive, the conditions for refinement 
must be specified.  In Enzo, the criteria for refinement can be set by the user to be
a combination of any or all of the following:  baryon or dark matter overdensity
threshold, minimum resolution of the local Jeans length, local density gradients,
local pressure gradients, local energy gradients, shocks, and cooling time.
A cell reaching
any or all of the user-specified criteria will then be flagged for refinement.  Once all 
cells of a given level have been flagged, rectangular solid boundaries are determined which 
minimally 
encompass them.  A refined grid patch is then introduced within each such bounding 
volume, and the results are interpolated to a higher level of resolution.

In Enzo, resolution of the equations being solved is adaptive in time as well as in
space.  The timestep in Enzo is satisfied on a level-by-level basis by finding the
largest timestep such that the Courant condition (and an analogous condition for 
the dark matter particles) is satisfied by every cell on that level.  All cells
on a given level are advanced using the same timestep.  Once a level $L$ has been
advanced in time $\Delta t_L$, all grids at level $L+1$ are 
advanced, using the same criteria for timestep calculations described above, until they
reach the same physical time as the grids at level $L$.  At this point grids at level
$L+1$ exchange baryon flux information with their parent grids, providing a more 
accurate solution on level $L$.  Cells at level $L+1$ are then examined to see 
if they should be refined or de-refined, and the entire grid hierarchy is rebuilt 
at that level (including all more highly refined levels).  The timestepping and 
hierarchy rebuilding process is repeated recursively on every level to the 
maximum existing grid level in the simulation.

Two different hydrodynamic methods are implemented in Enzo: the piecewise parabolic
method (PPM) \markcite{Woodward84}({Woodward} \& {Colella} 1984), which was extended to cosmology by 
\markcite{Bryan95}{Bryan} {et~al.} (1995), and the hydrodynamic method used in the ZEUS magnetohydrodynamics code
\markcite{stone92a,stone92b}({Stone} \& {Norman} 1992a, 1992b).  We direct the interested reader to the papers describing 
both of these methods for more information, and note that PPM is the preferred choice
of hydro method since it is higher-order-accurate and is based on a technique that 
does not require artificial viscosity, which smoothes shocks and can smear out 
features in the hydrodynamic flow.

The chemical and cooling properties of primordial (metal-free) gas are followed 
using the method of \markcite{abel97}{Abel} {et~al.} (1997) and \markcite{anninos97}{Anninos} {et~al.} (1997).  
This method follows the non-equilibrium evolution of a 
gas of primordial composition with 9 total species:  
$H$, $H^+$, $He$, $He^+$, $He^{++}$, $H^-$, $H_2^+$, $H_2$, and $e^-$.  The code 
also calculates 
radiative heating and cooling following atomic line excitation, recombination,
collisional excitation, free-free transitions, molecular line excitations, and Compton
scattering of the cosmic microwave background, as well as any of
approximately a dozen different models for a metagalactic UV background that heat
the gas via photoionization and photodissociation. 
The multispecies rate equations are solved out of
equilibrium to properly model situations where, e.g., the cooling time of the gas
is much shorter than the hydrogen recombination time.  
A total of 9 kinetic equations are solved, including 29 kinetic and radiative 
processes, for the 9 species mentioned above.  
The chemical reaction equation network is technically challenging to solve due to 
the huge range of reaction time scales involved -- the characteristic creation
and destruction time scales of the various species and reactions can differ by 
many orders of magnitude.  As a result, the set of rate equations is extremely 
stiff, and an explicit scheme for integration of the rate equations can be 
costly if small enough timestep are taken to keep the network
stable.  This makes an implicit scheme preferable for such a set of 
equations, and Enzo solves the rate equations using a method based on a backwards 
differencing formula (BDF) in order to provide a stable and accurate solution.
 
It is important to note the regime in which this chemistry model is valid.  According to 
\markcite{abel97}{Abel} {et~al.} (1997) and \markcite{anninos97}{Anninos} {et~al.} (1997), the reaction network is valid for temperatures
between $10^0 - 10^8$ K.  The original model discussed in these two references is only
valid up to n$_H \sim 10^4$~cm$^{-3}$.  However, addition of the 3-body H$_2$ formation
process allows correct modeling of the gas chemistry up to the point where the optically
thin cooling approximation begins to break down, at $\sim 10^{10}-10^{11}$~cm$^{-3}$.
Beyond this point, 
modifications to the cooling function that take into account the non-negligible
opacity of the gas to line radiation from molecular hydrogen must be made, as 
discussed by \markcite{ripamonti04}{Ripamonti} \& {Abel} (2004).  Even with these modifications, a more correct 
description of the cooling of gas of primordial composition at high densities will 
require some form of radiation transport, which will greatly 
increase the cost of the simulations.

\subsection{Effects of a gravitino WDM model}\label{wdmmodel}

As discussed previously, the primary effect of a warm dark matter particle is
suppression of the linear power spectrum at small scales.  There are currently two favored
warm dark matter candidates: gravitinos and sterile neutrinos.
We refer readers interested in a discussion of the particle physics motivation for 
these medium-mass particles,
as well as a discussion of their detailed properties, to work by 
\markcite{2001ApJ...556...93B}{Bode} {et~al.} (2001) and \markcite{abazajian05-2}{Abazajian} (2005b), and references therein.  
It is useful to note that the power spectrum for these warm dark matter 
candidates varies (see, e.g. \markcite{abazajian05-2}{Abazajian} (2005b)).  The sterile neutrino
candidate may have secondary effects which may significantly change the evolution 
of structure in the early universe (\markcite{biermann06}{Biermann} \& {Kusenko} (2006); see Section~\ref{discuss} for a discussion 
of this) so for simplicity we choose to use the gravitino warm dark matter 
of~\markcite{2001ApJ...556...93B}{Bode} {et~al.} (2001).  Due to this, we expect that application 
of our results to specific warm dark matter particle candidates may require some 
renormalization of the particle masses, though the results presented in this work
should be qualitatively correct for both warm dark matter candidates.

\markcite{2001ApJ...556...93B}{Bode} {et~al.} (2001) derive a formula for the power spectrum cutoff due to 
the existence of a gravitino warm dark matter particle.  They provide the following 
transfer function that models the smoothing of small-scale density perturbations:

\begin{equation} 
T_k^X = [1 + (\alpha k)^2]^{-5}
\label{wdm-tfctn}
\end{equation}

Where 
$\alpha = 0.05 (\Omega_{WDM}/0.4)^{0.15} (h/0.65)^{1.3} (keV/m_{WDM})^{1.15} (1.5/g_x)^{0.29}$ 
and k is in units of h Mpc$^{-1}$.  In this equation $\Omega_{WDM}$ is the contribution 
of the warm dark matter species to the energy density of the universe, in units of 
the critical density, m$_{WDM}$ is the gravitino mass in keV, $h$ is the Hubble 
constant in units of 100 km/s/Mpc, and $g_x$ is a parameter meant to represent the 
effective number of relativistic species present at decoupling, and is taken to 
be 1.5.  This transfer function is applied on top of the standard 
CDM transfer function, and imposes a strong rollover in the power spectrum at small 
scales, corresponding to a spatial smoothing scale of:

\begin{equation}
R_s \simeq 0.31 \left( \frac{\Omega_{WDM}}{0.3} \right)^{0.15} \left( \frac{h}{0.65} \right)^{1.3}  \left( \frac{keV}{m_{WDM}} \right)^{1.15} h^{-1} Mpc
\label{wdm-radius}
\end{equation}

This corresponds to the comoving half-wavelength of the mode at 
which the linear perturbation amplitude is suppressed by a factor of 2.  This 
smoothing
results in a characteristic mass scale below which structure forms by the top-down 
fragmentation of halos, rather than by the bottom-up hierarchical structure 
formation associated with the cold dark matter paradigm, which can be 
quantified as:

\begin{equation}
M_s = 10^{10} \left( \frac{\Omega_{WDM}}{0.3} \right)^{1.45}  
\left( \frac{h}{0.65} \right)^{3.9} \left( \frac{ keV }{ m_{WDM} } \right)^{3.45}   h^{-1} M_\odot
\label{wdm-radius}
\end{equation}

\begin{figure}
\epsscale{0.8}
\plotone{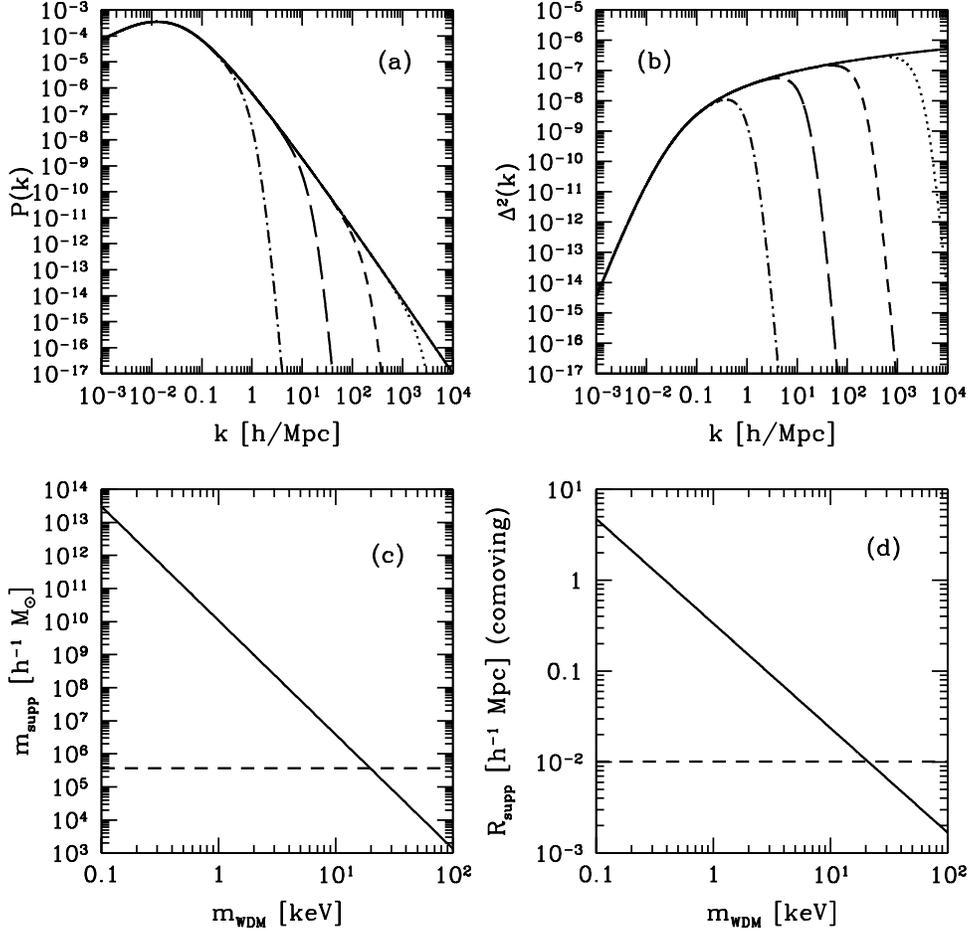}
\caption{
Plots showing various effects of a gravitino warm dark matter particle.
Panel (a):  The $z=0$ dark matter power spectrum $P(k)$ as a function
of wavenumber $k \equiv 2\pi/\lambda$ for a CDM cosmology and a range of
warm dark matter masses.  Panel (b):  The $z=0$ dimensionless power,
$\Delta^2(k) \sim k^3 P(k)$, versus k.  Panel (c):  The ``suppression
mass'' as a function of WDM particle mass.  Panel (d):  The spatial ``smoothing
scale'' as a function of WDM particle mass.  In plots (a) and (b) the solid line
corresponds to a CDM universe and the dotted, short dashed, long dashed and dot-short
dashed lines correspond to cosmologies with gravitino masses
of m$_{WDM} =$~100, 10, 1 and 0.1 keV, respectively.  In plots (c) and (d) the
horizontal dashed line
indicates the mass and radius corresponding to a halo of mass 
$4 \times 10^5$~M$_\odot$, which is approximately the mean halo mass from a set of
adaptive mesh refinement calculations of Population III star formation in a 
$\Lambda$CDM universe (O'Shea \&
Norman 2006, in preparation).
}
\label{fig.wdm-theory}
\end{figure}

Figure~\ref{fig.wdm-theory} contains several panels demonstrating the effects 
of the warm dark matter cosmology discussed above.  Panel (a) shows the 
cosmological power spectrum $P(k)$ at $z=0$ for a CDM cosmology as well as for 
when the WDM transfer function has been applied for several different
warm dark matter masses ranging from $0.1 - 100$ keV.  Panel (b) shows the 
dimensionless linear power $\Delta^2(k) \sim k^3 P(k)$ with the same 
particle masses.  Panel (c) shows the suppression mass as a function of 
radius, and panel (d) shows the comoving spatial smoothing scale.  In 
panels (c) and (d), the dashed line indicates the mass and radius corresponding 
to a halo of mass $4 \times 10^5$~M$_\odot$, which is roughly the mean halo mass 
for a suite of simulations of Population III star formation in a $\Lambda$CDM
universe (O'Shea \& Norman 2006, in preparation).  Note that this mass scale,
which corresponds to a gravitino mass of $\simeq$ 15 keV, is 
consistent with previous estimates, given some ambiguity in the definition of
the halo mass and collapse epoch and details of the initial simulation setup 
in different 
studies~\markcite{1997ApJ...474....1T,2001ApJ...548..509M,2002ApJ...564...23B,2003ApJ...592..645Y}({Tegmark} {et~al.} 1997; {Machacek}, {Bryan}, \&  {Abel} 2001; {Bromm} {et~al.} 2002; {Yoshida} {et~al.} 2003a).

\subsection{Simulation setup}\label{simsetup}

The simulations discussed in this paper are set up as follows.  A dark matter-only
calculation with $128^3$ particles in a three-dimensional simulation volume 
which is $0.3$~h$^{-1}$~Mpc (comoving) on a side
is set up at $z=99$ assuming a ``concordance''
cosmological model with no baryons:  $\Omega_m = \Omega_{DM} = 0.3$, 
$\Omega_b = 0.0$, $\Omega_\Lambda = 0.7$, $h=0.7$ (in units of 100 km/s/Mpc), 
$\sigma_8 = 0.9$, and using an Eisenstein \& Hu power spectrum \markcite{eishu99}({Eisenstein} \& {Hu} 1999)
with a spectral index of $n = 1$.  The cold dark matter cosmological model
is assumed.  This calculation is then evolved to 
$z=15$ using a maximum of four levels of adaptive mesh refinement, 
refining on a dark matter overdensity of 8.0.  At $z=15$, the 
Hop halo finding algorithm \markcite{eishut98}({Eisenstein} \& {Hut} 1998) is used to find the most massive 
halo in the simulation.

 At this point, we generate several sets of
initial conditions with the same large-scale structure by smoothing the CDM initial conditions
described above
with the warm dark matter transfer function described by Equation~\ref{wdm-tfctn},
assuming $\Omega_b = 0.04$, $\Omega_{WDM} = \Omega_{DM} = 0.26$, $g_x = 1.5$, and 
warm dark matter particle masses
of m$_{WDM} =$~10, 12.5, 15, 17.5, 20, 25, 30, 35, and 40~keV.  All other
cosmological parameters are identical.  An additional
calculation is performed with identical parameters but assuming the CDM
model (m$_{WDM} = \infty$).  The
initial conditions are generated
with both dark matter and baryons 
such that the Lagrangian volume in which the halo in the CDM case formed
is resolved at high spatial and mass resolution using a series of static 
nested grids, with a $128^3$ 
root grid and three static nested grids, for an overall effective root grid size of $1024^3$
cells.
The highest resolution grid  is $256^3$ grid cells, and corresponds
to a volume $75$~h$^{-1}$ comoving kpc on a side.
The dark matter particles in the highest
resolution grid are 1.81~h$^{-1}$~M$_\odot$ and the spatial resolution
of the highest resolution grid is 293~h$^{-1}$ parsecs (comoving). 
Previous work shows that this particle mass resolution is more than adequate 
to fully resolve the collapse of the halo (Abel, Bryan \& Norman, 2002; 
O'Shea \& Norman 2006, in preparation).

All simulations are performed using the adaptive
mesh cosmology code Enzo, which is described  in Section~\ref{enzocode}.  
The simulations are started at $z=99$ and allowed to evolve until the collapse
of the gas within the center of the most massive halo.  The 
equations of hydrodynamics
are solved using the PPM method with a dual energy formulation which is required 
to adequately resolve the thermal properties of gas in high-Mach flows.  The nonequilibrium 
chemical
evolution and optically thin radiative cooling of the primordial gas is 
modeled as described in Section~\ref{enzocode}, following 9 
separate species including molecular hydrogen (but excluding deuterium), with an initial
electron fraction of $1.2 \times 10^{-5}$ (which is roughly consistent with
\markcite{1968ApJ...153....1P}{Peebles} (1968)).  Note that the initial electron fraction in the 
calculation is relatively unimportant to the molecular hydrogen formation rates
in halo cores, as the electron fraction at the center of a given halo is 
controlled primarily
by mergers and the shock formed by accretion of gas onto the halo.  

Adaptive
mesh refinement is used such that cells are refined by factors of two along each 
axis, with a maximum of 22 total levels of refinement.  This corresponds to a 
maximum spatial resolution of 115~h$^{-1}$ astronomical units (comoving)
at the finest level of resolution, with an overall spatial dynamical range of
$5.37 \times 10^8$.  To avoid effects due to the finite size of the dark matter
particles, the dark matter density is smoothed on a comoving scale of $\sim 0.5$~pc
(which corresponds to $\simeq$ 0.03 proper pc at $z \simeq 18$).
This is reasonable because at that radius in all of our calculations the gravitational
potential is completely dominated by the baryons.

Grid cells are adaptively refined based upon several criteria:
baryon and dark matter overdensities in cells of 4.0 and 8.0, respectively, as well 
as criteria to ensure that the pressure jump and/or energy ratios between adjoining
cells never exceeds 5.0, that the cooling time in a given cell is always longer
than the sound crossing time of that cell, and that the Jeans length is always
resolved by at least 16 cells.  This last criterion guarantees that the Truelove
criterion \markcite{truelove97}({Truelove} {et~al.} 1997) is always resolved by a factor of four more cells 
in each dimension than is strictly necessary, ensuring that no artificial fragmentation 
will take place.

\section{Results}\label{results}

\subsection{General results}\label{results-general}

\begin{figure}
\plotone{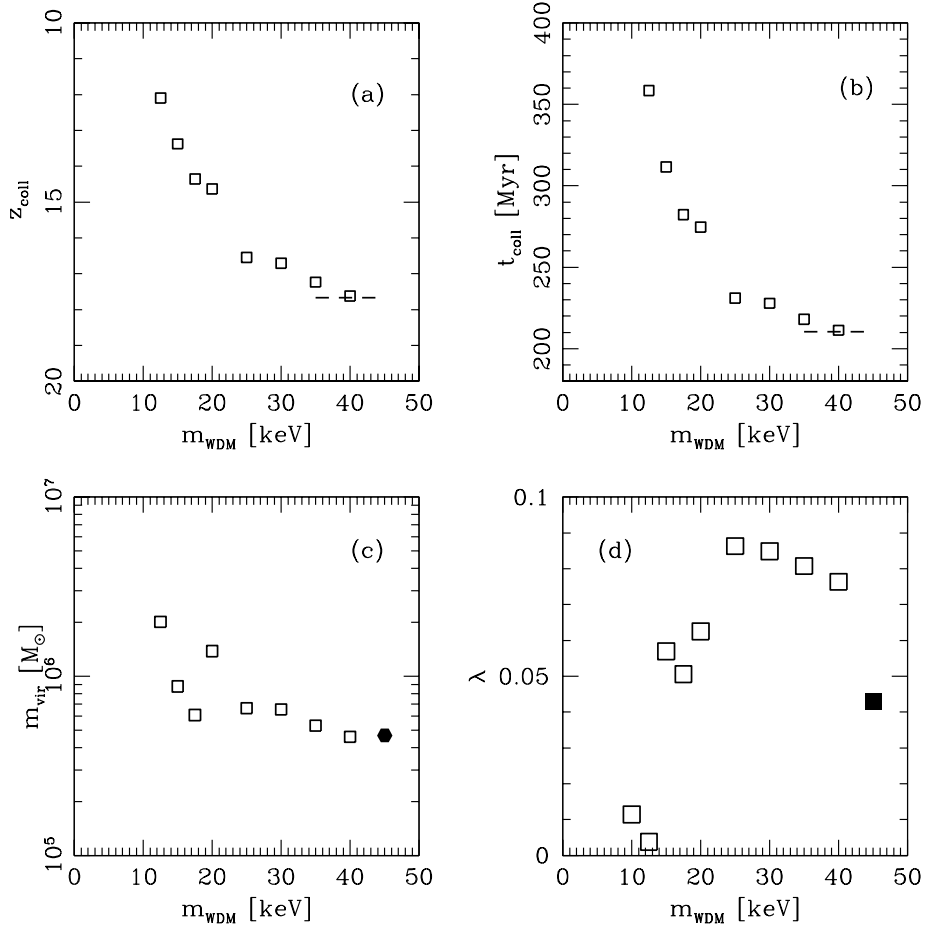}
\caption{
Dark matter halo properties as a function of WDM particle mass for several 
simulations with the same cosmological realization but different warm dark 
matter particle masses.  Panel (a):  WDM particle mass vs. collapse redshift 
of halo core.  Panel (b):  WDM particle mass vs. collapse time of halo core 
(measured in millions of years after the Big Bang).  Panel (c):  halo virial 
mass at collapse vs. WDM particle mass.  Panel (d):  halo dark matter
spin parameter at collapse vs. WDM particle mass.  In panels (a) and (b) 
the collapse redshift/time of the cold dark matter (CDM) simulation is 
shown by a horizontal dashed line.  In panel (c) the virial mass of
the halo in the CDM simulation is shown by a solid circle, while the 
WDM simulations are represented by open squares.  In panel (d) the spin
parameter of the halo in the CDM simulation is shown by a solid square, while
the WDM simulations are represented by open squares.
}
\label{fig.comp-mass-panel1}
\end{figure}

In this section we present the results of a comparison of all of the 
warm dark matter simulations, along with the cold dark matter ``control'' 
simulation.  Figure~\ref{fig.comp-mass-panel1} shows bulk properties of the 
halo in which the Population III protostar forms as a function of the
warm dark matter particle mass.  The top left and right panels plot the
WDM particle mass versus the redshift (left) and time after the Big Bang 
(right) at which the halo core collapses (where ``collapse'' in
this paper is defined as when the central baryon density of the halo reaches
$\sim 10^{10}$~cm$^{-3}$).
The CDM result is shown in both panels by a vertical dashed line.  Clearly,
 decreasing the warm dark matter particle mass
delays the formation of the protostar.  The 40 keV calculation forms
at essentially the same time as the CDM simulation, while collapse of the
halo core in the calculation assuming a 12.5 keV WDM particle mass is
delayed by approximately 130 million years compared to the CDM result.  
The simulation with a 10 keV
particle mass does not collapse by $z=10$ (the end of the simulation) and
is not shown here.  The delay of the halo collapse appears to be generally
smoothly varying as a function of WDM particle mass.

The bottom left panel of Figure~\ref{fig.comp-mass-panel1} shows the virial mass
of the halo (at the redshift of collapse) as a function of WDM particle mass.
A reduction in the WDM particle mass leads to an increase in the halo virial
mass, which is related to the delay in collapse of the halo core -- by the time
the halo core collapses in the lower particle mass simulations, the halo has
had time to accrete more mass.
The bottom right panel of Figure~\ref{fig.comp-mass-panel1} shows the
dark matter spin parameter of each halo (at the time of core collapse) as
a function of the gravitino particle mass.  It is clear that there is a relationship
between halo mass and spin parameter, but the overall cause and implications are 
unclear.  The gas spin parameter (not shown) displays similar behavior.

\begin{figure}
\plotone{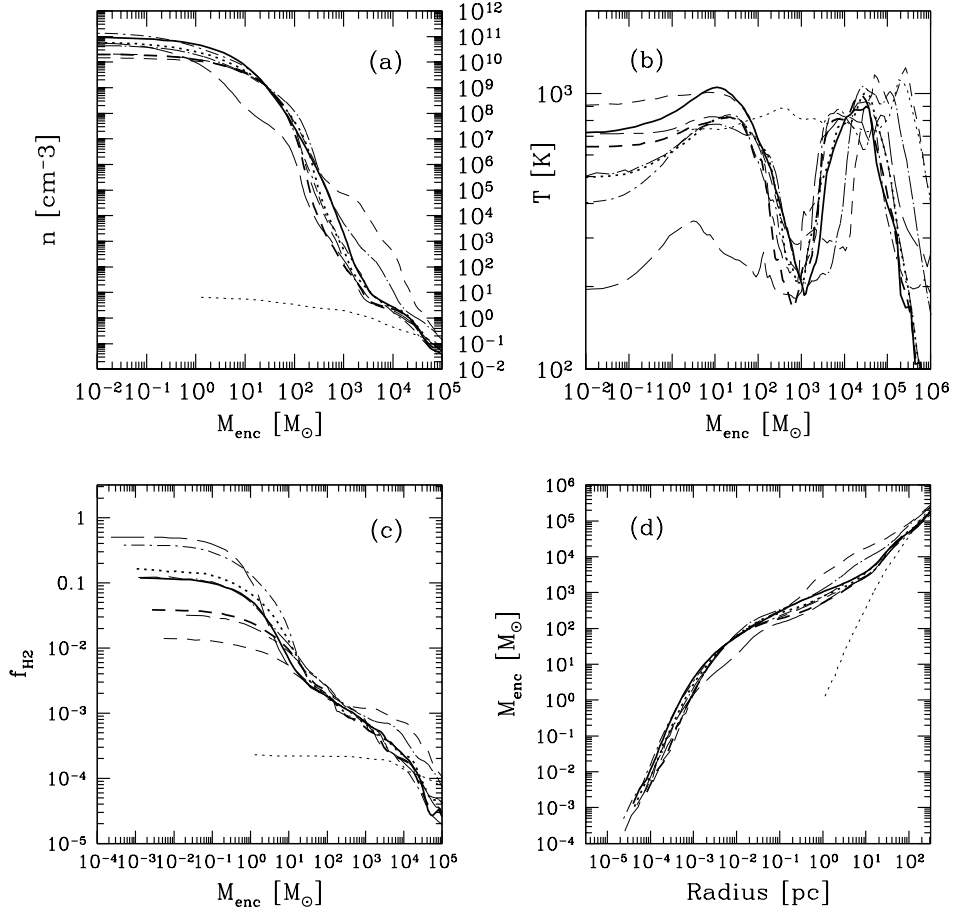}
\caption{
Mass-weighted, spherically-averaged baryon quantities for several simulations 
with the same cosmological realization but different warm dark matter particle 
masses.  Panel (a):  baryon number density as a function of enclosed baryon mass.  
Panel (b):  baryon temperature as a function of
enclosed baryon mass.  Panel (c): molecular hydrogen fraction as a function of enclosed
baryon mass.  Panel (d):  enclosed baryon mass as a function of radius.  Output times 
are chosen such that the peak baryon density in each simulation is approximately 
the same.  In each panel, the CDM simulation is represented by a thick solid line.
The 10, 12.5, 15, 17.5, 20, and 25 keV simulations are represented by thin
dotted, short-dashed, long-dashed, dot short-dashed, dot long-dashed, and 
short dash-long dashed lines, respectively.  The 30 and 35 keV calculations are 
represented by thick dotted and short-dashed lines, respectively.
}
\label{fig.comp-mass-panel2}
\end{figure}

\begin{figure}
\plotone{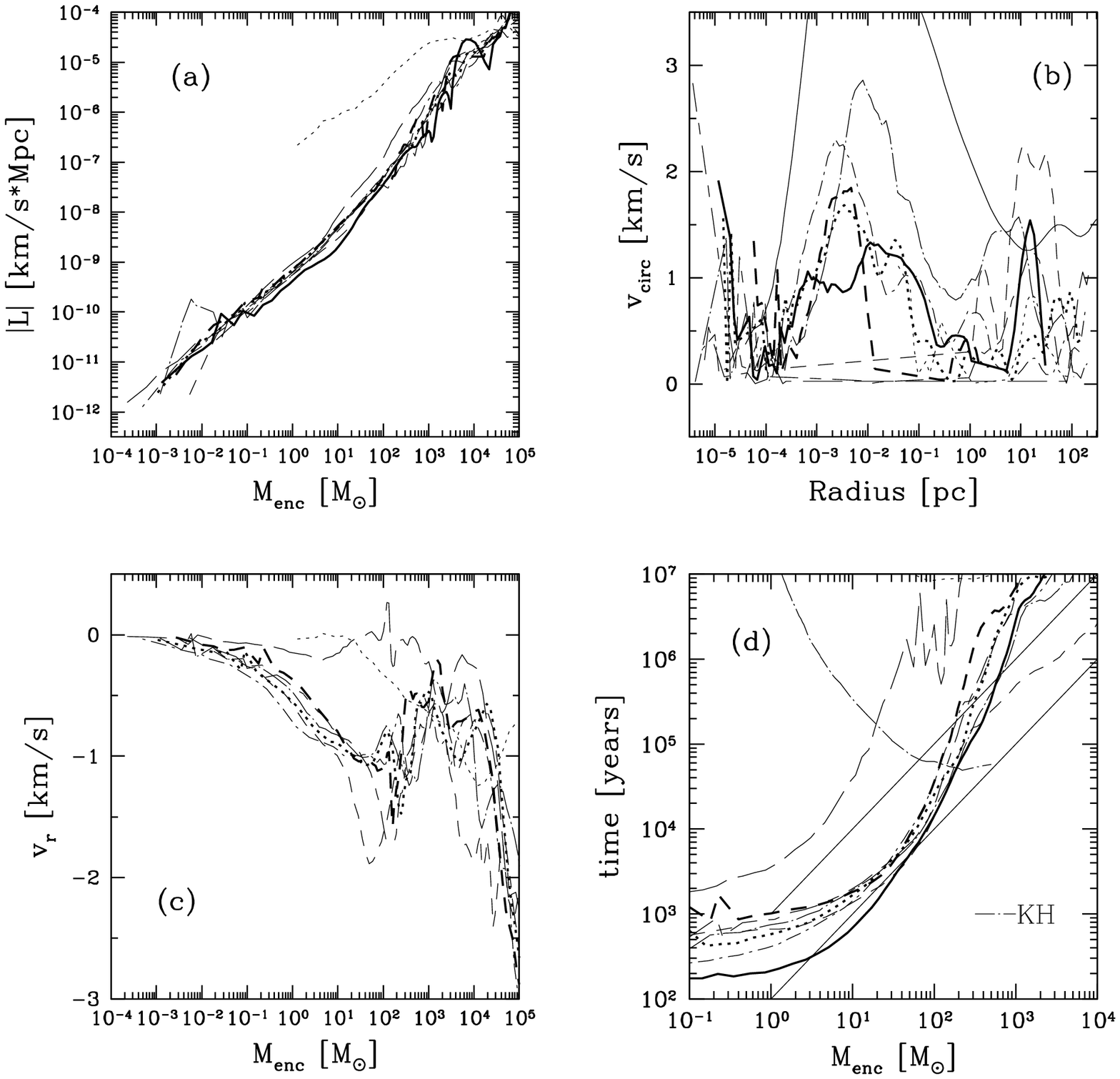}
\caption{
Mass-weighted baryon quantities for several simulations 
with the same cosmological realization but
different warm dark matter particle masses.  Panel (a): spherically-averaged
baryon angular momentum as
a function of enclosed baryon mass.  Panel (b):  cylindrically-averaged 
circular velocity a function of radius.  
Panel (c): spherically-averaged radial velocity as a function of enclosed baryon
mass.  Panel (d):  spherically-averaged accretion rate as a function of enclosed baryon mass.  
Output times are
chosen such that the peak baryon density in each simulation is approximately the same.
Line types are the same as in Figure~\ref{fig.comp-mass-panel2}.  The 
m$_{WDM} = 10$~keV case does not collapse by the end of the simulation and is 
shown at the last available output time in panels (a)-(c), and not shown in panel (d).
In Panel (d) the dot-long
dashed line which is roughly perpendicular to the others 
is the Kelvin-Helmholz time calculated from Population III stellar 
properties from \markcite{2002A&A...382...28S}{Schaerer} (2002) and the
 upper and lower diagonal thin
solid lines correspond to constant accretion rates of $10^{-3}$ 
and $10^{-2}$~M$_\odot$/yr, respectively.
}
\label{fig.comp-mass-panel3}
\end{figure}

Figures~\ref{fig.comp-mass-panel2} and~\ref{fig.comp-mass-panel3} show 
several spherically-averaged, mass-weighted radial profiles of baryon 
quantities as a function of radius or enclosed baryon mass of the simulations.  
All profiles are taken at a constant point in the evolution of the protostellar 
cloud (when the proper central baryon number density is n~$\sim 10^{10}$~cm$^{-3}$) 
rather than at a constant point in time, since the halos collapse over a wide 
range of redshifts.  Figure~\ref{fig.comp-mass-panel2} shows the proper baryon 
number 
density as a function of enclosed baryon mass (Panel (a)), baryon temperature
as a function of enclosed baryon mass (Panel (b)), molecular hydrogen fraction as a 
function of enclosed baryon mass (Panel (c)), and enclosed baryon mass as a function of
radius (Panel (d)).  As expected, the number density profiles of all of the 
simulations are very similar over the entire range of WDM (and CDM) particle 
masses.  This is a result of the cooling properties of the gas.  In the 
absence of any other sources of energy (such as magnetic fields or cosmic 
rays), self-gravitating gas in pressure equilibrium tends toward an isothermal 
(r$^{-2}$) density profile, which is seen here.  The simulation assuming a  
WDM particle mass of 10 keV does not collapse by the time the simulation is 
stopped at $z=10$, and the density profile at the last output time is shown.  
The plot of enclosed mass as a function of radius shows a strong similarity 
between the different calculations as well, which is to be expected as it is 
essentially another way of viewing the number density plot.  The plots of 
temperature and molecular hydrogen fraction vs. enclosed mass show a 
significant amount of scatter.  Ignoring the 10 keV case, the overall spread 
in temperature in the core of the halo is a factor of $\sim 3$ and the spread 
in molecular hydrogen fraction is roughly 1.5 orders of magnitude.  
It is interesting to note that the CDM ``control'' simulation has one of the 
higher core temperatures and a median $H_2$ fraction.  There is no apparent 
relationship between the warm dark matter particle mass and final baryon 
quantities.

Figure~\ref{fig.comp-mass-panel3} shows the specific angular momentum as a 
function of enclosed baryon mass (Panel (a)), circular velocity as a function of 
radius (Panel (b)), radial velocity as a function of enclosed baryon mass (Panel (c)), 
and accretion time as a function of enclosed baryon mass (Panel (d)).  The angular 
momentum distributions are somewhat similar for all of the calculations 
(disregarding the 10 keV case since it does not collapse), though there is
a great deal of scatter at large radii, which is also shown in the scatter
in spin parameter in Figure~\ref{fig.comp-mass-panel1} (d).  The 
circular velocities are similar as well, though also with a noticeable scatter.
  The Keplerian orbital velocity for the CDM (least massive) halo is 
plotted in this panel (upper thin black line), and all of the simulations 
display circular velocities that are significantly below this velocity.  The 
plot of radial velocity as a function of enclosed baryon mass shows that the CDM 
simulation has the greatest infall velocity at the output time in question, 
which corresponds to the largest accretion rate overall (as shown in the plot 
of accretion time vs. enclosed baryon mass).  The rest of the calculations have 
similar infall velocities and accretion rates, except for the 15 keV model, 
which has a much lower overall infall velocity and accretion rate.  The reason
for this is unclear.  The overall accretion rates for the WDM calculations 
are slightly less than that of the CDM calculation, suggesting that the final 
stellar masses may be slightly lower, if one applies the same techniques for 
estimating the stellar mass as used by \markcite{ABN02}{Abel} {et~al.} (2002).  However, it has been
argued that the magnitude of the accretion rate onto a primordial protostar
at the time of core collapse may not be directly related to the final mass
\markcite{2003ApJ...589..677O}({Omukai} \& {Palla} 2003), so the effect of the change of accretion rates is unclear.

\begin{figure}
\plotone{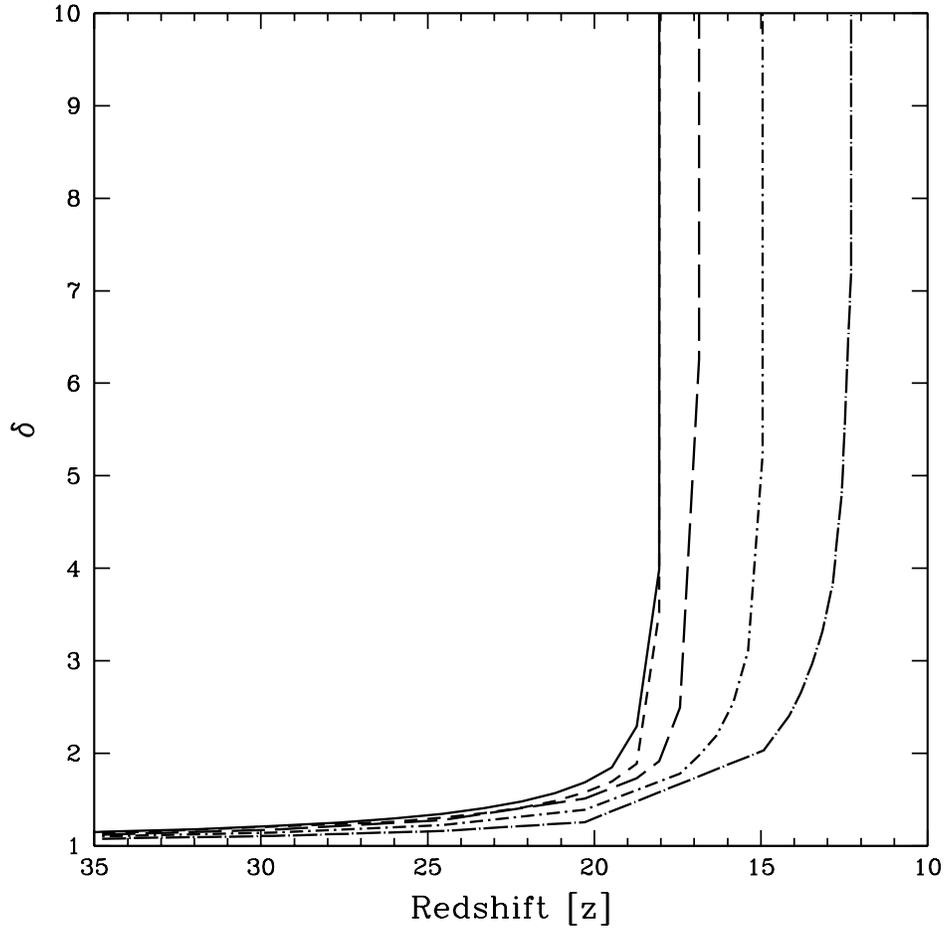}
\caption{
Clumping factor as a function of redshift for representative warm dark matter models.
The CDM calculation is represented by a black solid line.  The m$_{WDM} = $~12.5, 20, 30 
and 40 keV models are represented by dot-long-dashed, dot-short-dashed, long-dashed, 
and short-dashed lines, respectively.
}
\label{fig.comp-mass-panel4}
\end{figure}

Figure~\ref{fig.comp-mass-panel4} shows the evolution of clumping factor as a function
of redshift for the CDM case (solid black line) and four representative gravitino
particle masses: 12.5, 20, 30 and 40 keV.  s in the other quantities, the 40 keV 
run is almost indistinguishable from CDM, while lowering the gravitino particle mass
decreases the clumping factor for a given redshift.  This may have implications for
reionization, though the simulation volume is too small to draw meaningful 
statistical conclusions (as discussed in Section~\ref{discuss}).

\subsection{Comparison of two representative models}\label{results-rep}

\begin{figure}
\plotone{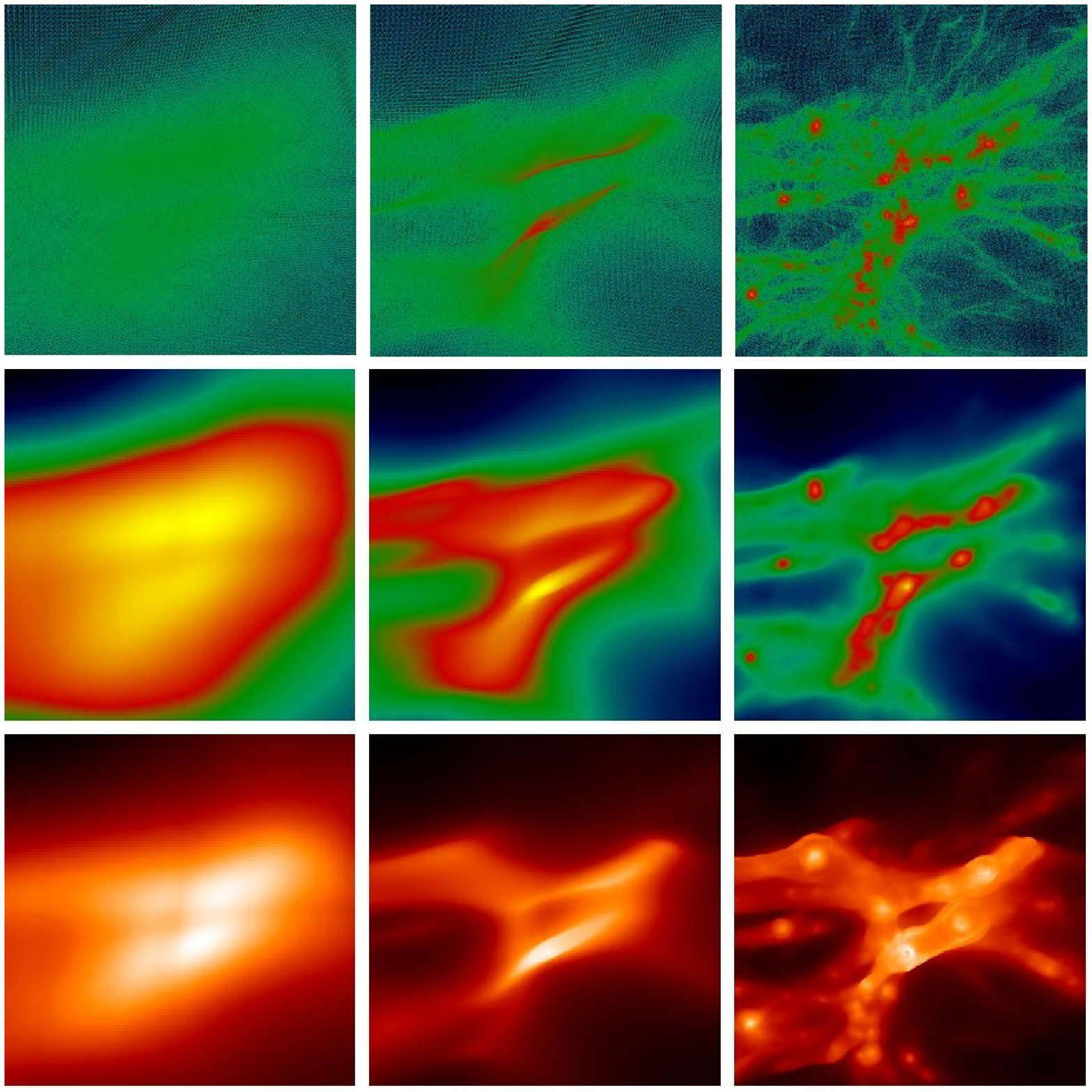}
\caption{
Mass-weighted projections of dark matter density, baryon density and baryon 
temperature for 3 simulations with the same cosmological realization and a range 
of warm dark matter (WDM) particle masses at $z=20.38$.  The field in each 
calculation is the same, though the color tables are relative for each panel in 
order to highlight density differences.  Left column:  m$_{WDM} = 12.5$~keV.  
Center column:  m$_{WDM} = 25$~keV.  Right column: Cold dark matter realization 
(corresponds to m$_{WDM} \rightarrow \infty$).  Top row:  projected dark matter 
density.  Middle row:  projected baryon density.  Bottom row:  projected,
mass-weighted baryon 
temperature.  The spatial scale is $\sim 1.3$~kpc (proper) in each volume.
}
\label{fig.comp-mass-image1}
\end{figure}

In this section we compare the evolution of two representative warm dark matter 
simulations.  We choose the calculations with WDM particle masses of $12.5$ and 
$25$ keV, as these are examples of simulations where the warm dark matter
``suppression mass'' is somewhat above (12.5 keV) or below (25 keV) the mean
CDM Population III halo mass of $\simeq 4 \times 10^5$~M$_\odot$, suggesting that these halos will form via top-down
fragmentation or hierarchical mergers, respectively, and a comparison of the 
primordial protostellar core's properties in each case will be illuminating.  
Figure~\ref{fig.comp-mass-image1} shows mass-weighted projections of dark matter 
density, baryon density, and baryon temperature at $z=20.38$
for the two representative WDM calculations and a CDM calculation of the same 
cosmological realization.  All panels show a volume that is $\sim 300$~pc 
(proper) across and are centered on the point in space where the first Population 
III protostar will form.  There is a significant difference between the calculations 
at a fixed point in time -- the cold dark matter calculation (right column) 
shows a great deal of clumpy dark matter structure, including knots along 
the cosmological filaments and even dark matter halos in void regions, 
with corresponding variety in the baryon density and temperature
plots.  The 25 keV calculation shows the effects of smoothing - two halos are 
forming, but there are no halos in the voids, and no substructure around the 
halos that form.  This is reflected in the baryon temperature and density plots, 
where the accretion shocks onto the filaments show little small-scale structure 
and the gas is quite smooth.  The 12.5 keV calculation is an even more striking 
example of the effects of suppressing the power spectrum at small scales -- 
though an overdensity in the dark matter is apparent, no halos are visible at 
this redshift and there is no smaller scale structure whatsoever.
This warm dark matter particle mass corresponds to a smoothing scale of a few 
times $10^6$~M$_\odot$, below which it has been speculated that top-down 
fragmentation takes place (as observed by \markcite{2003MNRAS.345.1285K}{Knebe} {et~al.} (2003), though
this has been contested by \markcite{2003Ap&SS.284..341G}{G{\"o}tz} \& {Sommer-Larsen} (2003)), and is 
roughly equivalent to the mass of the coalescing halo shown in this image.

\begin{figure}
\plotone{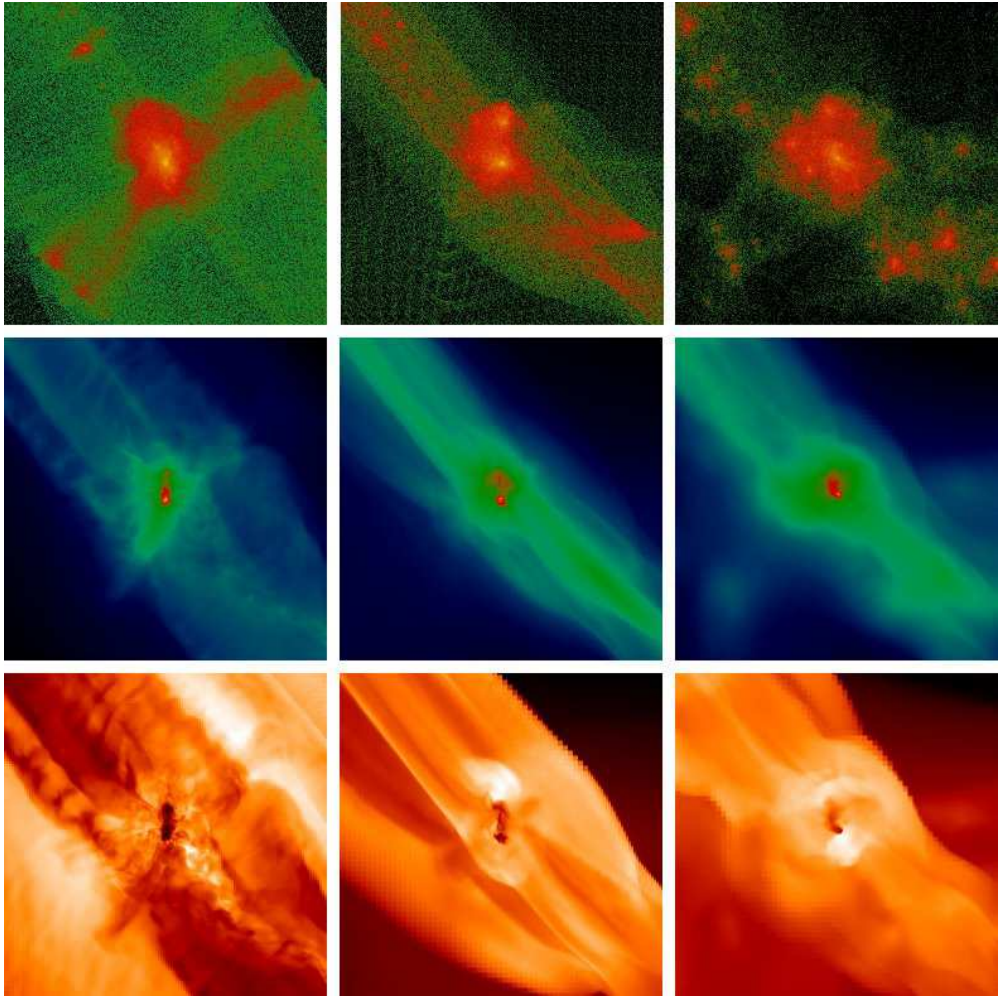}
\caption{
Mass-weighted projections of dark matter density, baryon density and baryon 
temperature for 3 simulations with the same cosmological realization and a 
range of warm dark matter (WDM) particle masses at the redshift at which the 
Population III protostar collapses in each simulation.  The comoving size of 
the projected volume in each calculation is the same, though the color tables 
are relative for each panel in order to highlight density differences.
Left column:  m$_{WDM} = 12.5$~keV, $z_{coll} = 12.09$.  Center column:  
m$_{WDM} = 25$~keV, $z_{coll} = 16.54$.  Right column: Cold dark matter 
realization (corresponds to m$_{WDM} \rightarrow \infty$), $z_{coll} = 18.001$.
Top row:  projected dark matter density.  Middle row:  projected baryon density.
Bottom row:  projected baryon temperature.  The spatial scale is $\sim 300$~pc 
(proper) for the CDM and m$_{WDM} = 25$~keV WDM simulations and $\sim 450$~pc 
(proper) for the m$_{WDM} = 12.5$~keV WDM simulation; the comoving scales 
are the same in each panel.
}
\label{fig.comp-mass-image2}
\end{figure}

Figure~\ref{fig.comp-mass-image2} shows the same quantities and spatial volume 
as Figure~\ref{fig.comp-mass-image1}, though instead of the outputs all being 
at the same point in time, they are at the time when the halo core collapses in 
each simulation.  This corresponds to $z=18.001$ for the CDM calculation, 
$z=16.54$ for the WDM calculation with m$_{WDM} = 25$~keV, and $z=12.09$ for the 
WDM calculation with m$_{WDM} = 12.5$~keV.  At the time of collapse the 12.5 keV 
calculation has formed a halo which is more massive than the CDM halo by a 
factor of $\sim 5$ (and collapses approximately 130 million years later).  Very 
little substructure is evident in the projected dark matter distribution of the 
12.5 keV calculation.  Some is apparent in the 25 keV run, but not nearly as 
much as in the CDM calculation.  As predicted by \markcite{2001ApJ...556...93B}{Bode} {et~al.} (2001), the warm dark 
matter calculations have suppressed substructure and satellite halos, and it
is apparent that the halo in the 12.5 keV calculation forms by top-down 
fragmentation of a filament rather than hierarchical merging of smaller halos.
This top-down fragmentation results in a great deal of turbulent gas motion --
the results of this can be seen in the projected baryon density and temperature
plots in the left column of Figure~\ref{fig.comp-mass-image2}, and are discussed
in more detail in Section~\ref{results-topdown}.

\begin{figure}
\plottwo{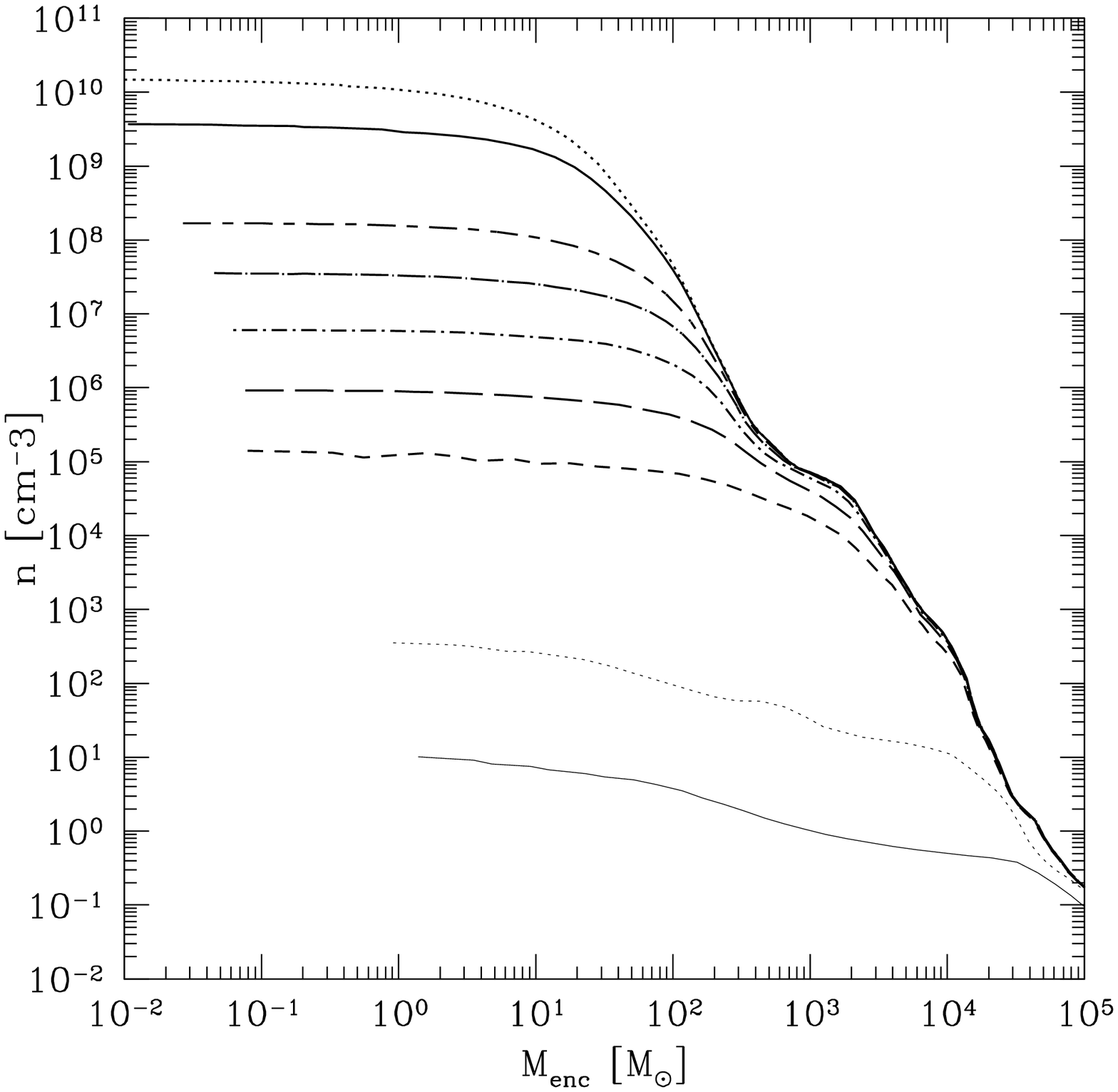}{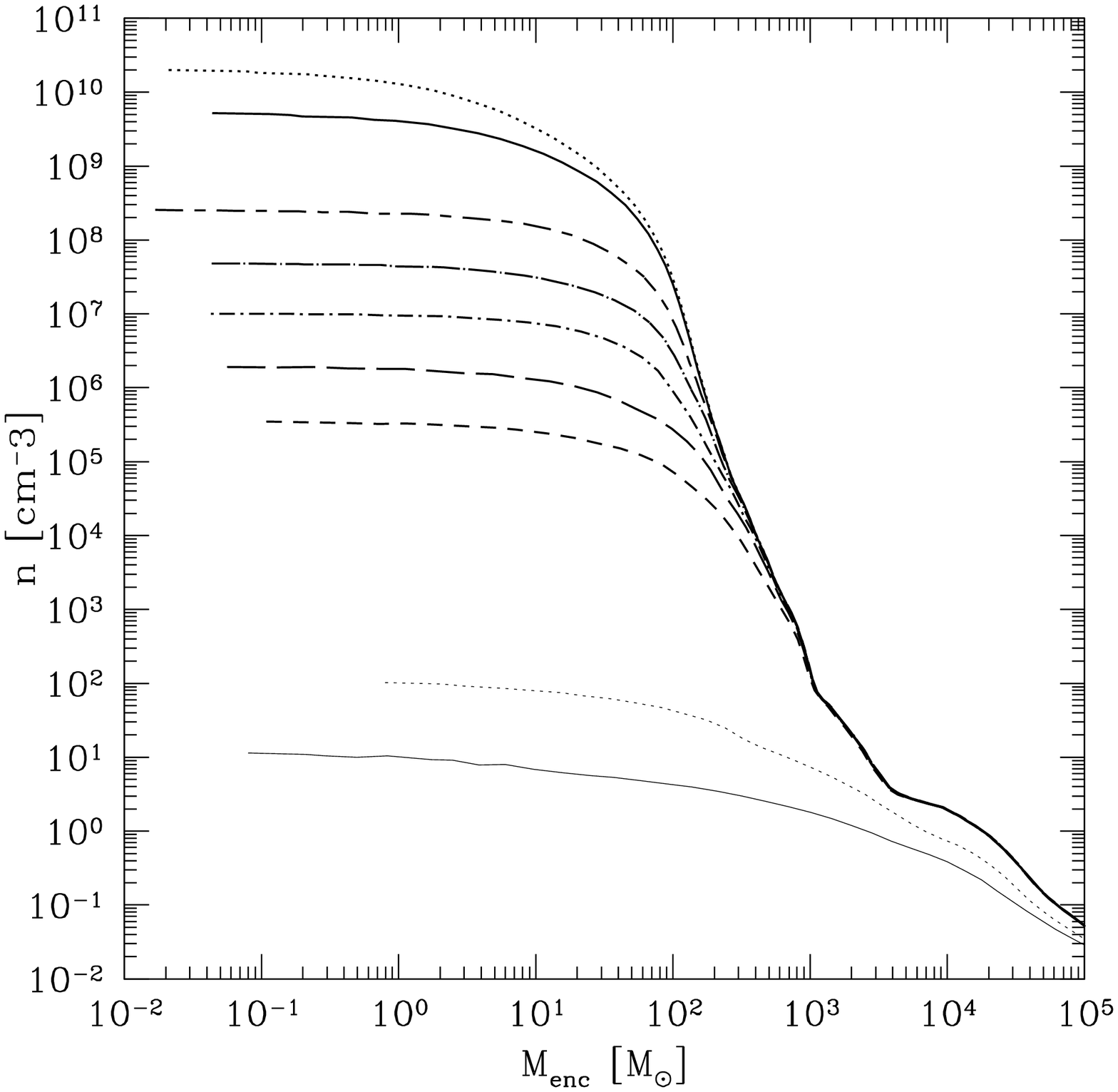}
\caption{
Evolution of spherically-averaged, mass-weighted baryon number density 
as a function of enclosed mass in halos with two representative warm dark 
matter simulations.  The cosmological realization is the same for each 
calculation, and output times are chosen such that the baryon densities 
are approximately the same.  Left column:  simulation with 
m$_{WDM} = 12.5$~keV.  Right column:  simulation with m$_{WDM} = 25$~keV.  
Lines for the m$_{WDM} = 12.5$~keV ($25$~keV) simulations as follows.  
Thin solid line:  $t = 319$ Myr/$z=13.163$ ($t = 204$ Myr/$z=18.05$).  
Thin dotted line: $3.12 \times 10^7$ years later ($1.04 \times 10^7$ years later).  
Thick short-dashed line: $8.15 \times 10^6$ years later ($1.04 \times 10^6$ years later).  
Thick long-dashed line: $98,345$ years later ($5.73 \times 10^6$ years later).
Thick dot-short-dashed line: $2.86 \times 10^5$ years later ($2.63 \times 10^5$ years later).
Thick dot-long-dashed line: $1.25 \times 10^5$ years later ($82,433$ years later).
Thick short dashed-long dashed line: $45,152$ years later ($38,738$ years later).
Thick solid line: $22,697$ years later ($24,865$ years later).
Thick dotted line: $2691$ years later ($3332$ years later).
}
\label{fig.comp-evol-panel1}
\end{figure}

\begin{figure}
\plotone{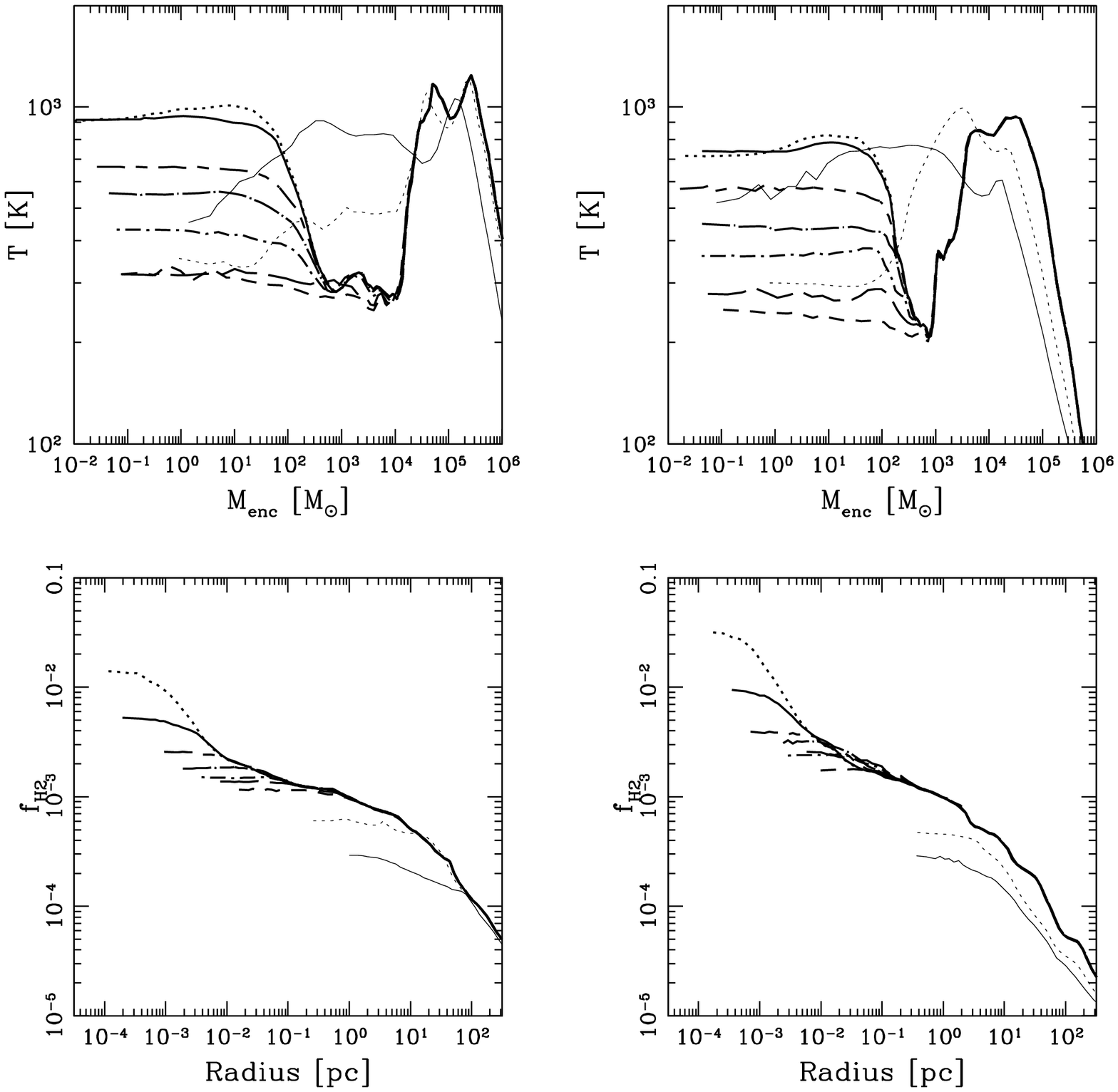}
\caption{
Evolution of spherically-averaged, mass-weighted baryon temperature (top row) 
and molecular hydrogen fraction (bottom row) as a function of enclosed mass
 in halos with two representative warm dark matter simulations.  The 
cosmological realization is the same for each calculation, and output times 
are chosen such that the baryon densities are approximately the same.  Left 
column:  simulation with m$_{WDM} = 12.5$~keV.  Right column:  simulation 
with m$_{WDM} = 25$~keV.  The lines are at the same times as in 
Figure~\ref{fig.comp-evol-panel1}.
}
\label{fig.comp-evol-panel2}
\end{figure}

\begin{figure}
\plotone{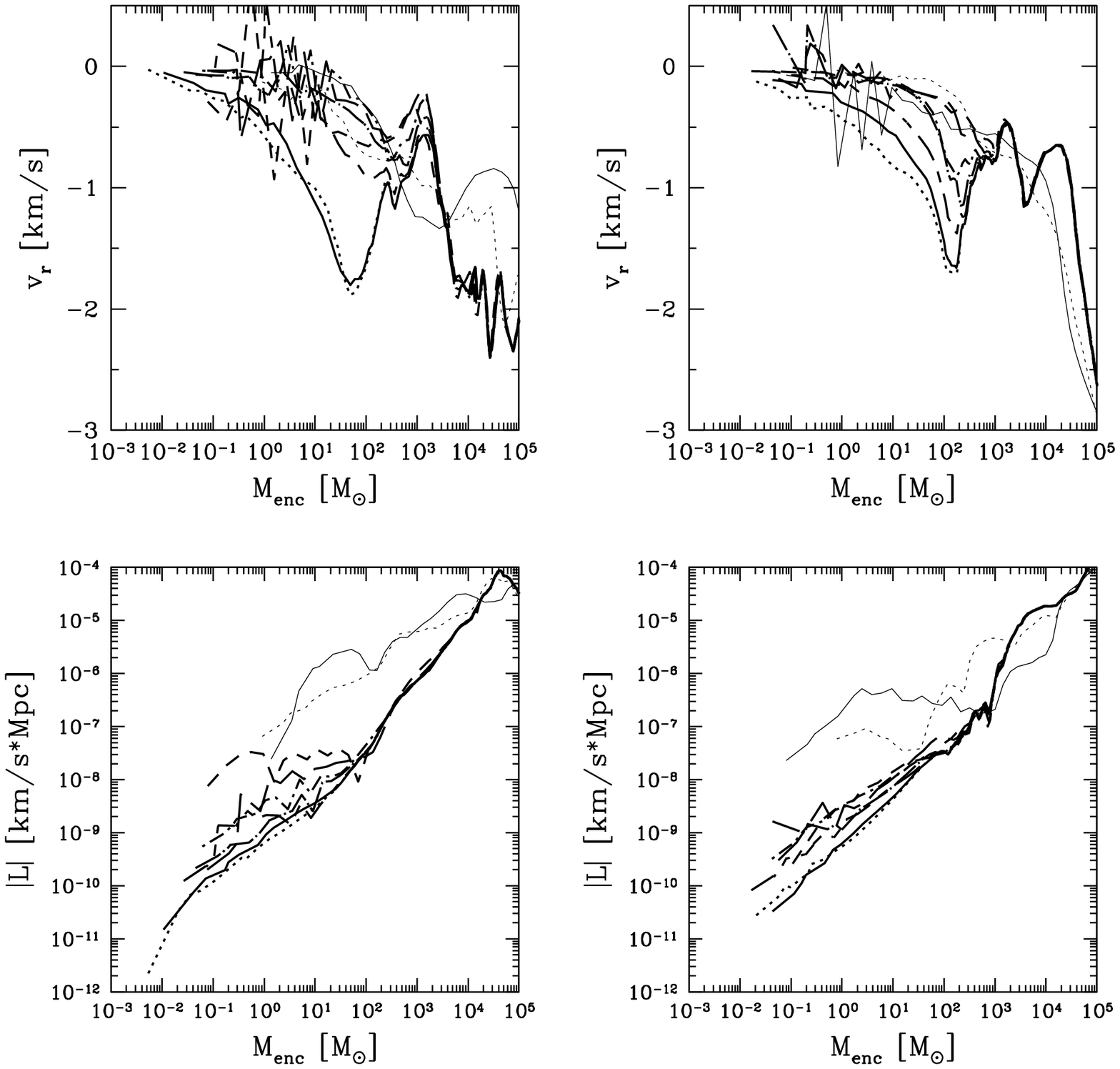}
\caption{
Evolution of spherically-averaged, mass-weighted baryon properties in halos 
with two representative warm dark matter simulations.  The cosmological 
realization is the same for each calculation, and output times are chosen
such that the baryon densities are approximately the same.  Left column:  
simulation with m$_{WDM} = 12.5$~keV.  Right column:  simulation with 
m$_{WDM} = 25$~keV.  Top row:  baryon radial velocity as a function of 
enclosed baryon mass (velocity is positive away from the center of the 
halo).  Bottom row:  baryon angular momentum as a function of enclosed 
mass.  The lines are at the same times as in Figure~\ref{fig.comp-evol-panel1}.
}
\label{fig.comp-evol-panel3}
\end{figure}

Figures~\ref{fig.comp-evol-panel1} through~\ref{fig.comp-evol-panel3} show
the time evolution of several spherically averaged, mass-weighted radial 
quantities for the two representative warm dark matter calculations.  The 
plots are chosen such that the central densities of the collapsing halo 
core are matched between the two calculations.

Figure~\ref{fig.comp-evol-panel1} shows the evolution of number density as 
a function of enclosed mass for the 12.5 keV and 25 keV WDM calculations.  
The lowest-density line corresponds to $z=13.16$ for the 12.5 keV run and 
$z=18.05$ for the 25 keV calculation.  Intriguingly, it takes the 12.5 keV 
calculation roughly $4 \times 10^7$ years to advance from a core
number density of  n~$\sim 10$~cm$^{-3}$ (proper) 
to a core baryon number 
density of n~$\sim 10^5$~cm$^{-3}$, while the 25 keV calculation only 
requires $\sim 2 \times 10^7$ years to evolve over the same range.  However, once 
the calculations reach $\simeq 10^5$~cm$^{-3}$ they take similar amounts 
of time to evolve to the highest number density shown. As discussed in 
previous sections, this reflects the fact that the halo evolution on small 
scales is controlled by the chemistry and cooling properties of the 
primordial gas, which is unaffected by the dark matter properties.  The slower 
evolution of the gas in the m$_{WDM} = 12.5$~keV simulation at low densities can be
attributed to the top-down formation of the halo -- the halo's potential
well is shallower overall (due to its more extended dark matter profile), 
leading to a shallower baryon density profile at the halo core and thus longer 
chemical reaction times and slower core evolution.

Figure~\ref{fig.comp-evol-panel2} shows the evolution of baryon temperature 
and molecular hydrogen fraction as a function of enclosed mass for the two warm
dark matter simulations.  The overall temperature evolution is very similar 
between the two calculations, though the calculation with m$_{WDM} = 12.5$~keV 
ends up with a slightly lower molecular hydrogen fraction and slightly higher 
central temperature.  The evolution of radial infall velocity and angular 
momentum as a function of enclosed mass (shown in 
Figure~\ref{fig.comp-evol-panel3}) is also quite similar between the two 
calculations.  As shown in Figure~\ref{fig.comp-mass-panel3}, the final 
accretion rates are also essentially the same.

The purpose of this section was to demonstrate that the evolution of the 
halo collapse, and the resulting protostar, is quite similar for two simulations 
with significantly different warm dark matter particle masses.  The large-scale 
structure evolves somewhat differently in these two cases -- the halo that forms 
in the 12.5 keV calculation is approximately the same mass scale as the 
suppression mass, meaning that this dark matter halo is roughly the smallest 
object that can directly form at that mass scale.  The halo that forms in the 
25 keV WDM model is significantly larger than the suppression mass, implying 
that it formed out of the merging of smaller objects.  Despite this, the final
protostellar properties are similar, which is due to the collapse dynamics 
being controlled at small scales primarily by the cooling 
properties of the primordial gas rather than by the dynamics of the large scale 
cosmological structure.

\subsection{Top-down formation of a Population III star-forming halo}\label{results-topdown}

Figure~\ref{fig.comp-mass-image1} (discussed in Section~\ref{results-rep}) shows evidence for 
significant turbulent gas motion in the halo which forms a Population III star 
in a m$_{wdm} = 12.5$~keV universe.  The gas which appears to be experiencing this turbulent 
motion is aligned with the filament out of which the cosmological minihalo coalesced, suggesting
some sort of large-scale motion is the cause of the turbulence.  This sort of structure
is not seen in 
our standard CDM calculation, nor in the m$_{wdm} = 25$~keV run.  The ``smoothing mass''
(the mass corresponding to the spatial scale at which the
power spectrum is suppressed by a factor of two compared to CDM) of the 12.5 keV 
calculation is a few times $10^6$~M$_\odot$, which is approximately 
the virial mass of the halo that eventually forms, whereas the same mass scale for the 25
keV calculation is $\simeq 10^5$~M$_\odot$, and is zero for CDM (by definition).  This
suggests that the halo in which the Population III star forms in the m$_{WDM}$ = 12.5 keV
calculation is the smallest object which can form, and should collapse as a single entity
rather than being built up through hierarchical merging.  How does this affect the 
properties of the baryons within the halo?

\begin{figure}
\plotone{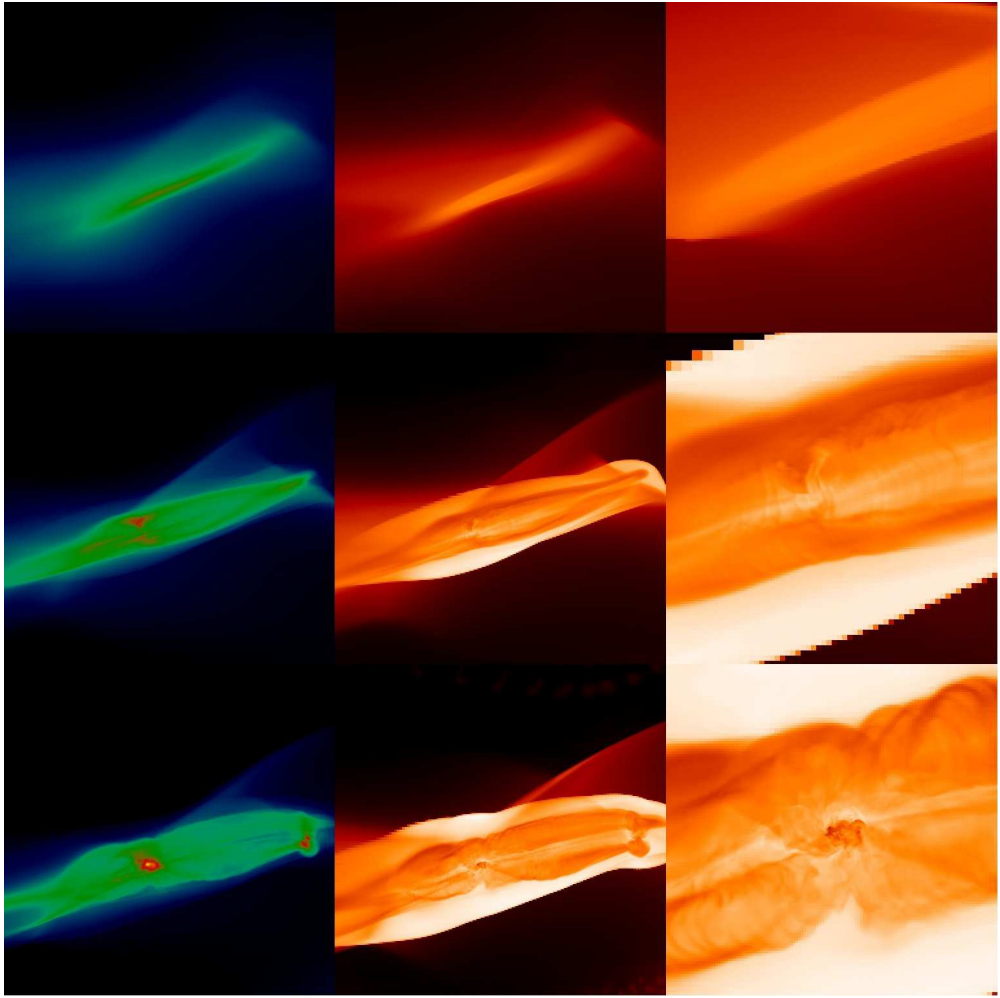}
\caption{
Images of baryon density and temperature as a function of redshift for the 
m$_{WDM} = 12.5$~keV simulation.  Left column:  projected baryon density.
Central and right columns: projected, mass-weighted baryon temperature.
Top row: quantities at $z=15.838$.  Center row: quantities at $z=13.480$
(62.4 Myrs after top row).  Bottom row:  quantities at $z=12.091$ (112.7 Myr
after top row).
The two left columns correspond to a proper spatial scale of $\simeq 1.5$~kpc
at $z=16$, while the right column is a zoom into the region where the Population
III protostar will form, and has a proper spatial scale of $\simeq 350$~pc at
$z=16$.  All projections are along the same axis of the simulation.  Image colors
in a given column correspond to the same baryon quantity in all panels.
}
\label{fig.topdown-panel1}
\end{figure}

Figure~\ref{fig.topdown-panel1} shows images of projected baryon density and temperature
taken at three different times during the collapse of this halo.  The top row shows the halo at a 
point in time shortly after it has become a distinct object.  In contrast
to a CDM Population III minihalo, the halo structure is greatly elongated and it appears 
that a caustic of some sort is beginning to form.  Examination of the baryon temperatures
show that gas has not yet been shocked to high values.  The center row shows the halo at
a later time ($\sim 6 \times 10^7$ years later), when the baryons have collapsed further.
The collapse occurs along the minor axis of the halo, and has heated the gas to relatively
high temperatures.  A cylindrical shock structure can be clearly seen at both large and 
small scales (center and right columns), and turbulent structures can be seen in regions
of high baryon density.  Due to the cylindrical collapse of the halo, these structures are
generally aligned with the filament.  The bottom row shows the halo at the time when the
Population III protostellar core reaches central baryon densities of $\sim 10^{10}$~cm$^{-3}$,
approximately $5 \times 10^7$ years after the previous row of images.  Turbulent structures
aligned with the filament are clearly visible, as is the shocked gas in the halo.

\begin{figure}
\plotone{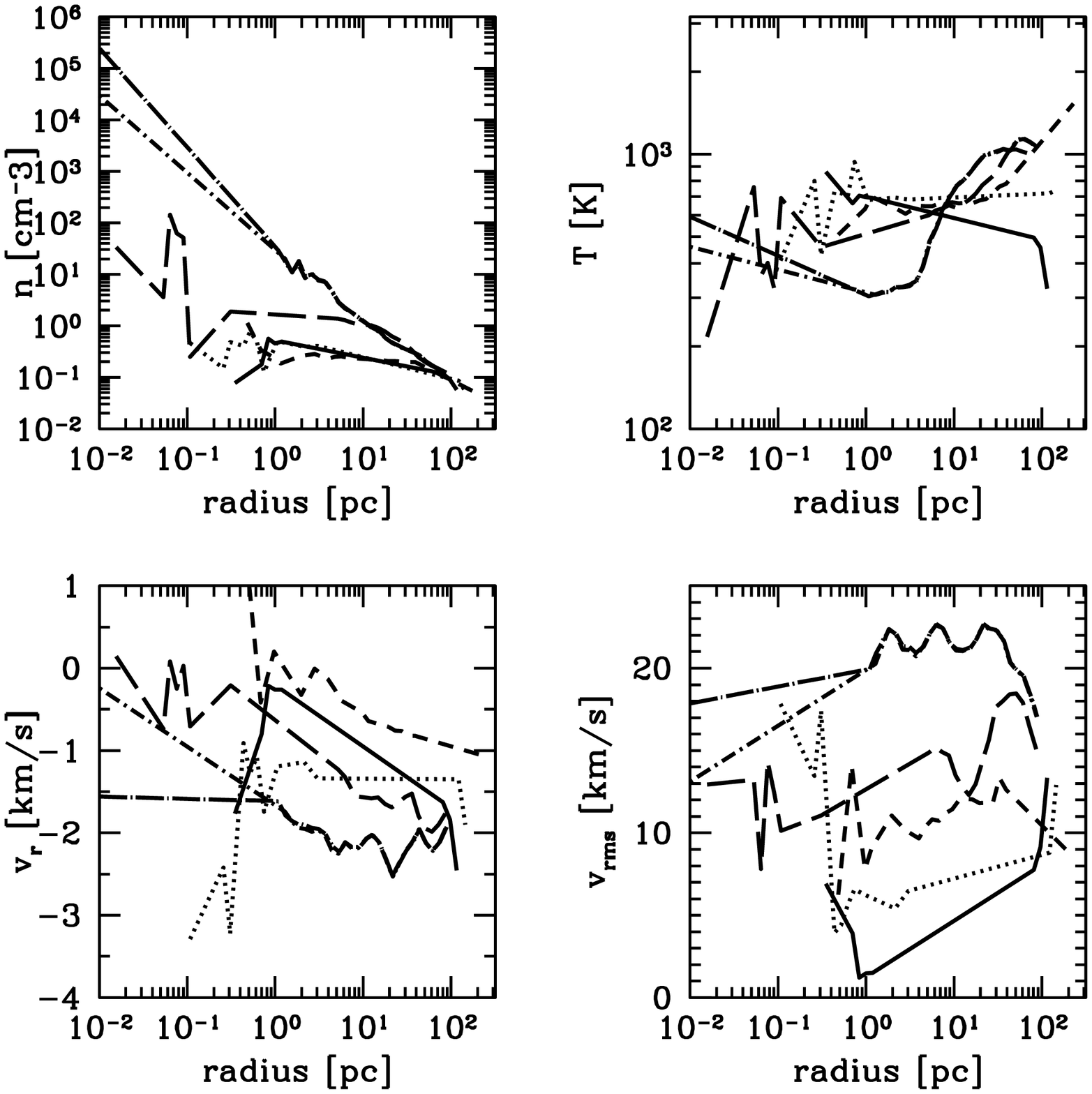}
\caption{
Cylindrically-averaged, mass-weighted radial profiles of baryon quantities 
during filament collapse in the m$_{WDM} = 12.5$~keV simulation.  Top
left panel:  baryon number density as a function of radius.  Top right
panel:  baryon temperature as a function of radius.  Bottom left 
panel:  baryon radial velocity as a function of radius.  Bottom
right panel:  baryon 3D RMS velocity as a function of radius.  Lines
in each panel correspond to identical times.  Solid line: $z=15.838$.
Dotted line: $z=14.950$.  Short-dashed line: $z=13.48$.  Long-dashed
line: $z=12.308$.  Dot-short-dashed line: $z=12.092$.  Dot-long-dashed
line: $z=12.091$.  At each time the center of the cylinder is the
maximum point of baryon density in the collapsing halo.
}
\label{fig.topdown-panel2}
\end{figure}

Figure~\ref{fig.topdown-panel2} shows the evolution of the turbulent structures in a 
more quantitative fashion.  This figure shows the time evolution of the 
cylindrically-averged baryon 
number density, temperature, radial velocity, and 3D RMS velocity over the range of times shown
in Figure~\ref{fig.topdown-panel1}.  
The cylindrical average is calculated with respect to the center
of the cosmological filament at each timestep shown, with the cylinder aligned with
the filament, and examing all gas with a proper density $\ga 10^{-2}$~cm$^{-3}$.
The halo is initially collapsing with a 
strong radial velocity and some non-radial motion (as can be seen by the 3D
RMS velocity).  Over time, as the halo collapses and gas shocks, the disordered
(i.e. turbulent) motion of the gas increases, as can be seen by the large
increase of gas RMS velocity with respect to the radial velocity.  Examination
of the temperature evolution of the gas shows an object which was initially 
generally quite cool, but was heated at large radii and late times
by the same large-scale shocks that created the turbulent gas motions.

\section{Discussion}\label{discuss}

There are several effects which we have neglected in these calculations which 
may affect our results.  The box size of the calculations presented in this 
paper is relatively small (300 kpc/h comoving).  \markcite{2004ApJ...609..474B}{Barkana} \& {Loeb} (2004)
show that finite box sizes can have significant effects on statistical properties 
of halo formation, resulting in an overall bias towards undersampling of the mass
function and late halo formation times.  Though their results were explicitly for
a CDM cosmology, they are also relevant for warm dark matter cosmologies.
However, our calculations are 
not intended to provide a statistically accurate sampling of the mass function 
of dark matter halos.  Rather, the intent is to gain a qualitative feel for the 
effects of a  warm dark matter power spectrum on a single halo.  An 
obvious extension of this work would be to simulate a much larger box size 
so that cosmic variance and possible halo environmental effects are considered 
(as discussed in \markcite{2004ApJ...609..474B}{Barkana} \& {Loeb} (2004)), and then simulate multiple halos 
within these boxes for a range of $\Lambda$WDM models, all compared to a 
$\Lambda$CDM ``control'' calculation.  A suite of simulations such as these is 
computationally costly, but not completely infeasible, and would lend 
statistical weight to the qualitative results presented in this work.

Another important consideration is the generalization of cosmological initial conditions.
\markcite{2003Ap&SS.284..341G}{G{\"o}tz} \& {Sommer-Larsen} (2003) show that grid-based warm dark matter initial conditions
can result in unphysical halos along filaments, giving the appearance of
``beads on a string,'' and producing an unreliable halo satellite mass
function.  They suggest using ``glass'' initial conditions \markcite{1996clss.conf..349W}({White} 1996)
to alleviate these problems.  While we do not use a glass distribution to
initialize our calculations, we are interested only in the most massive halo
in the simulation, not the overall or satellite halo distributions, and this effect
is unimportant for our result.  Similarly, \markcite{2006astro.ph..1233H}{Heitmann} {et~al.} (2006) explore
the evolution of the halo mass function at high redshifts, and conclude that
the overall mass function is suppressed in simulations whose initial conditions
are generated at too late of a redshift.  They suggest that a starting
simulation redshift of $z \simeq 500$ would be appropriate for the small boxes used in 
the calculations discussed in this paper.  This is impractical for simulations containing
baryonic physics, and the magnitude of error involved in simulations starting later is
not quantified.  Since our focus is only on the most massive halo to form in the
simulation volume, the magnitude of the bias in our results by starting at $z=100$ 
is unclear.

In this paper, we consider the effects of a gravitino warm dark matter power 
spectrum, as given by \markcite{2001ApJ...556...93B}{Bode} {et~al.} (2001).  The other viable warm 
dark matter candidate, sterile neutrinos, will produce a different 
power spectrum at small scales for a given particle mass, which will alter our 
results~\markcite{abazajian05-2}({Abazajian} 2005b).  However, the two WDM candidates both have similar
behavior (exponential suppression of power at small scales) so our result
may have some applicability to the sterile neutrino, with some rescaling of masses.
However, sterile neutrinos 
may have secondary effects which cloud the primary issues being examined in
this work:  \markcite{biermann06}{Biermann} \& {Kusenko} (2006) suggest that decaying sterile neutrinos in 
the early universe may produce a relatively high level of 
global ionization compared to the standard case ($f_e \sim 10^{-2}$ instead
of $\sim 10^{-4}$), which would serve to significantly increase molecular 
hydrogen production and possibly make Population III star formation more 
effective.  A gravitino model avoids these secondary effects, and allows us
to concentrate on more general issues.  A further improvement upon 
this work would be to use power spectra for the sterile neutrino 
candidate, and to include their secondary effects.

There are other examples of particles that may cause suppression of
small-scale power, such as superweakly interacting particles (``SuperWIMPs'')
\markcite{2005PhRvL..95r1301C}({Cembranos} {et~al.} 2005) or composite dark matter \markcite{2005astro.ph.11796K}({Khlopov} 2006).  
It is claimed that these particles may also
provide solutions to the small-scale issues with structure formation. 
These claims are testable in cosmological simulations, and would be an obvious 
extension to the results presented in this paper.

Once the baryons in all of the warm dark matter simulations reach significantly 
high densities (n$_b \sim 10^5$~cm$^{-3}$), the halo core collapses nearly 
identically.  This is because, when the core of the halo effectively decouples 
from the cosmological halo (i.e. t$_{dyn}$, t$_{s} \ll$ t$_H$, where t$_{dyn}$, 
t$_{s}$ and t$_H$ are the dynamical and sound crossing times of the halo core 
and the Hubble time, respectively) the evolution of the core is driven by gas 
chemistry and cooling physics, which is not affected by the cosmological model.  
This is seen in calculations of Population III star formation in a $\Lambda$CDM 
universe, where a wide range of halo masses and formation times are sampled with 
 similar results (O'Shea and Norman 2006, in preparation).

Current observations do not probe structure formation in the regime considered 
in this paper.  The most ambitious ground- and space-based telescopes which are 
currently being planned or are under construction will not be able to directly
observe individual Population III stars, even if they are quite massive, due 
to their extremely small angular size and large cosmological distance.  This 
suggests  that the most appropriate way to probe early structure formation may 
be by studying reionization.  Future CMB experiments, such as the PLANCK 
mission\footnote{http://planck.esa.int/}, 
as well as observations of redshifted 21 cm line emission will show how and when 
the universe was reionized, and may give us clues as to which cosmological model 
is more correct~\markcite{2003ApJ...583...24K}({Kaplinghat} {et~al.} 2003).  Reionization may be complex 
(e.g. \markcite{2003ApJ...591...12C}{Cen} (2003)), and these observations may need more advanced 
simulations to disentangle the effects of Population III and later structures.  
This has been partially done numerically by \markcite{2003MNRAS.344..607S}{Sokasian} {et~al.} (2003) and 
\markcite{2003ApJ...591L...1Y}{Yoshida} {et~al.} (2003b) but needs to be expanded to include a larger range 
of possible dark matter (e.g. linear power spectrum) models.  It must be noted 
that all of our discussion of constraints on early universe structure formation 
by examining an early epoch reionization are predicated upon the assumption that 
this reionization was due to stars, rather than, e.g., sterile neutrino 
decay~\markcite{2004ApJ...600...26H}({Hansen} \& {Haiman} 2004).

Warm dark matter models are somewhat analogous to CDM calculations with a soft
UV background, in that both cause an overall delay in collapse of the halo core
and result in halos with a somewhat larger virial mass (corresponding to the 
later collapse time).  This is due to different physical reasons, of course.  One
striking difference in the warm dark matter calculations is that for WDM
particle masses below $\simeq 15$~keV, the suppression mass is actually at the 
mass of the halo in which the primordial protostar forms (at a few times
$10^5$~M$_\odot$) so a different paradigm for structure formation occurs:  the
halos at this scale may form by top-down fragmentation of larger objects rather
than bottom-up formation via hierarchical mergers.  Remarkably, this does not
appear to strongly impact the Population III initial mass function.

In principle, examination of the delay in halo collapse could allow us to suggest
a new constraint on the gravitino particle mass by comparing to measurements of
the reionization history of the universe.  However, even the lowest gravitino mass
in a simulation where the halo collapsed, M$_{WDM} =$~12.5 keV, is still consistent
with the WMAP Year 3 polarization measurement.  Given the small box size and
the problems inherent in extrapolating the reionization 
history of the universe from a single object, this should not be taken as any
sort of useful limit.  In order to provide a more plausible limit to the gravitino
particle mass, simulations of much larger cosmological volumes must be run and the 
inferred reionization histories compared to future CMB and 21 cm observations.

\section{Summary of results}\label{summary}

In this paper we have performed a series of simulations of structure formation 
in the early universe in an attempt to provide more stringent constraints on a 
possible warm dark matter particle mass and to examine how the suppression of
small-scale power affects the formation of protostellar cores.  We use an 
identical cosmological realization, but apply smoothing to the initial 
conditions according to the WDM transfer function given by 
\markcite{2001ApJ...556...93B}{Bode} {et~al.} (2001), for a range of warm dark matter particle masses 
which have not yet been observationally ruled out.  Our principal conclusions 
are as follows:

1.  We find that, for a wide range of warm dark matter particle masses, the 
main effect of the smoothing of small-scale power is to delay the collapse 
of high-density gas at the center of the most massive halo in the simulation 
and, as a result, an increase in the virial mass of this halo at the onset of 
baryon collapse.  Both of these effects become more pronounced as the warm 
dark matter particle mass becomes smaller.

2.  A cosmology using a warm dark matter particle mass of $\simeq 40$~keV is
effectively indistinguishable from the cold dark matter case, making this a reasonable
upper limit of gravitino masses that have any significant effect
on large scale structure (though providing no useful particle physics
constraints on the mass of such a particle).

3.  There is remarkably little scatter in the final properties of the 
primordial protostellar core which forms at the center of the halo, possibly 
due to the overall low rate of halo mergers which is a result of the warm dark 
matter power spectrum.  

4.  The detailed evolution of the collapsing halo core in two representative 
warm dark matter cosmologies is described.  At relatively low densities 
(n$_b \la 10^5$~cm$^{-3}$) the evolution of the two calculations is similar,
but the calculation with a lower particle mass takes a much longer amount of 
time to reach the runaway collapse phase.  Once the gas in the center of the 
halo reaches higher densities (n$_b \ga 10^5$~cm$^{-3}$), the overall evolution 
is essentially identical in the two calculations, implying that the mass of 
the warm dark matter particle should not affect the Population III initial
mass function, assuming that the particle mass is small enough to allow
prototypical Population III halos to still form.

\acknowledgments{BWO would like to thank Kevork Abazajian,
Tom Abel, Greg Bryan, Salman Habib, Katrin Heitmann, and Matthew Turk 
for useful discussions.  We would like to thank an anonymous referee
who made several suggestions which have improved the quality of this 
manuscript.
This work was supported in part by NASA
grant NAG5-12140 and NSF grant AST-0307690. 
BWO has been funded in part
under the auspices of the U.S.\ Dept.\ of Energy, and supported by its
contract W-7405-ENG-36 to Los Alamos National Laboratory.  The
simulations were performed at SDSC and NCSA with computing time provided by 
NRAC allocation MCA98N020.  
}


\begin{thebibliography}{}

\bibitem[{Abazajian} 2005a]{abazajian05}
{Abazajian}, K. 2005a, PRD, submitted; astro-ph/0512631

\bibitem[{Abazajian} 2005b]{abazajian05-2}
---. 2005b, PRD, submitted; astro-ph/0511630

\bibitem[{Abazajian}, {Fuller}, \&  {Tucker} 2001]{2001ApJ...562..593A}
{Abazajian}, K., {Fuller}, G.~M., \& {Tucker}, W.~H. 2001, \apj, 562, 593

\bibitem[{Abel}, {Anninos}, {Zhang}, \&  {Norman} 1997]{abel97}
{Abel}, T., {Anninos}, P., {Zhang}, Y., \& {Norman}, M.~L. 1997, New Astronomy,  2, 181

\bibitem[{Abel}, {Bryan}, \& {Norman} 2002]{ABN02}
{Abel}, T., {Bryan}, G.~L., \& {Norman}, M.~L. 2002, Science, 295, 93

\bibitem[{Anninos}, {Zhang}, {Abel}, \&  {Norman} 1997]{anninos97}
{Anninos}, P., {Zhang}, Y., {Abel}, T., \& {Norman}, M.~L. 1997, New Astronomy,  2, 209

\bibitem[{Bahcall}, {Ostriker}, {Perlmutter}, \&  {Steinhardt} 1999]{bahcall99}
{Bahcall}, N.~A., {Ostriker}, J.~P., {Perlmutter}, S., \& {Steinhardt}, P.~J.  1999, Science, 284, 1481

\bibitem[{Bardeen}, {Bond}, {Kaiser}, \&  {Szalay} 1986]{1986ApJ...304...15B}
{Bardeen}, J.~M., {Bond}, J.~R., {Kaiser}, N., \& {Szalay}, A.~S. 1986, \apj,  304, 15

\bibitem[{Barkana}, {Haiman}, \&  {Ostriker} 2001]{2001ApJ...558..482B}
{Barkana}, R., {Haiman}, Z., \& {Ostriker}, J.~P. 2001, \apj, 558, 482

\bibitem[{Barkana} \& {Loeb} 2004]{2004ApJ...609..474B}
{Barkana}, R. \& {Loeb}, A. 2004, \apj, 609, 474

\bibitem[{Berger} \& {Colella} 1989]{Berger89}
{Berger}, M.~J. \& {Colella}, P. 1989, J. Comp. Phys., 82, 64

\bibitem[{Biermann} \& {Kusenko} 2006]{biermann06}
{Biermann}, P. \& {Kusenko}, A. 2006, astro-ph/0601004

\bibitem[{Bode}, {Ostriker}, \&  {Turok} 2001]{2001ApJ...556...93B}
{Bode}, P., {Ostriker}, J.~P., \& {Turok}, N. 2001, \apj, 556, 93

\bibitem[{Bromm}, {Coppi}, \&  {Larson} 2002]{2002ApJ...564...23B}
{Bromm}, V., {Coppi}, P.~S., \& {Larson}, R.~B. 2002, \apj, 564, 23

\bibitem[{Bryan} \& {Norman} 1997a]{bryan97}
{Bryan}, G. \& {Norman}, M. 1997a, 12th Kingston Meeting on  Theoretical Astrophysics, proceedings of meeting held in Halifax; Nova  Scotia; Canada October 17-19; 1996 (ASP Conference Series \# 123), ed.  D.~Clarke. \& M.~Fall

\bibitem[{Bryan} \& {Norman} 1997b]{bryan99}
---. 1997b, Workshop on Structured Adaptive Mesh Refinement Grid  Methods, ed. N.~Chrisochoides (IMA Volumes in Mathematics No. 117)

\bibitem[{Bryan}, {Norman}, {Stone}, {Cen}, \&  {Ostriker} 1995]{Bryan95}
{Bryan}, G.~L., {Norman}, M.~L., {Stone}, J.~M., {Cen}, R., \& {Ostriker},  J.~P. 1995, Comp. Phys. Comm, 89, 149

\bibitem[{Cembranos}, {Feng}, {Rajaraman}, \&  {Takayama} 2005]{2005PhRvL..95r1301C}
{Cembranos}, J.~A., {Feng}, J.~L., {Rajaraman}, A., \& {Takayama}, F. 2005,  Physical Review Letters, 95, 181301

\bibitem[{Cen} 2003]{2003ApJ...591...12C}
{Cen}, R. 2003, \apj, 591, 12

\bibitem[{Chiba} 2002]{2002ApJ...565...17C}
{Chiba}, M. 2002, \apj, 565, 17

\bibitem[{Colombi}, {Dodelson}, \&  {Widrow} 1996]{1996ApJ...458....1C}
{Colombi}, S., {Dodelson}, S., \& {Widrow}, L.~M. 1996, \apj, 458, 1

\bibitem[{Dalal} \& {Kochanek} 2002]{2002ApJ...572...25D}
{Dalal}, N. \& {Kochanek}, C.~S. 2002, \apj, 572, 25

\bibitem[{Dalcanton} \& {Hogan} 2001]{2001ApJ...561...35D}
{Dalcanton}, J.~J. \& {Hogan}, C.~J. 2001, \apj, 561, 35

\bibitem[{Efstathiou}, {Davis}, {White}, \&  {Frenk} 1985]{Efstathiou85}
{Efstathiou}, G., {Davis}, M., {White}, S.~D.~M., \& {Frenk}, C.~S. 1985,  \apjs, 57, 241

\bibitem[{Eisenstein} \& {Hu} 1999]{eishu99}
{Eisenstein}, D.~J. \& {Hu}, W. 1999, \apj, 511, 5

\bibitem[{Eisenstein} \& {Hut} 1998]{eishut98}
{Eisenstein}, D.~J. \& {Hut}, P. 1998, \apj, 498, 137

\bibitem[{Ghigna}, {Moore}, {Governato}, {Lake},  {Quinn}, \& {Stadel} 2000]{2000ApJ...544..616G}
{Ghigna}, S., {Moore}, B., {Governato}, F., {Lake}, G., {Quinn}, T., \&  {Stadel}, J. 2000, \apj, 544, 616

\bibitem[{G{\"o}tz} \& {Sommer-Larsen} 2003]{2003Ap&SS.284..341G}
{G{\"o}tz}, M. \& {Sommer-Larsen}, J. 2003, \apss, 284, 341

\bibitem[{Governato}, {Mayer}, {Wadsley}, {Gardner},  {Willman}, {Hayashi}, {Quinn}, {Stadel}, \& {Lake} 2004]{2004ApJ...607..688G}
{Governato}, F., {Mayer}, L., {Wadsley}, J., {Gardner}, J.~P., {Willman}, B.,  {Hayashi}, E., {Quinn}, T., {Stadel}, J., {et al.} 2004, \apj, 607, 688

\bibitem[{Hansen} \& {Haiman} 2004]{2004ApJ...600...26H}
{Hansen}, S.~H. \& {Haiman}, Z. 2004, \apj, 600, 26

\bibitem[{Heitmann}, {Lukic}, {Habib}, \&  {Ricker} 2006]{2006astro.ph..1233H}
{Heitmann}, K., {Lukic}, Z., {Habib}, S., \& {Ricker}, P.~M. 2006, ArXiv  Astrophysics e-prints

\bibitem[{Hockney} \& {Eastwood} 1988]{Hockney88}
{Hockney}, R.~W. \& {Eastwood}, J.~W. 1988, Computer Simulation Using Particles  (Institute of Physics Publishing)

\bibitem[{Kaplinghat}, {Chu}, {Haiman}, {Holder},  {Knox}, \& {Skordis} 2003]{2003ApJ...583...24K}
{Kaplinghat}, M., {Chu}, M., {Haiman}, Z., {Holder}, G.~P., {Knox}, L., \&  {Skordis}, C. 2003, \apj, 583, 24

\bibitem[{Khlopov} 2006]{2005astro.ph.11796K}
{Khlopov}, M.~Y. 2006, Pisma Zh.Eksp.Teor.Fiz., 83, 3

\bibitem[{Klypin}, {Kravtsov}, {Valenzuela}, \&  {Prada} 1999]{1999ApJ...522...82K}
{Klypin}, A., {Kravtsov}, A.~V., {Valenzuela}, O., \& {Prada}, F. 1999, \apj,  522, 82

\bibitem[{Knebe}, {Devriendt}, {Gibson}, \&  {Silk} 2003]{2003MNRAS.345.1285K}
{Knebe}, A., {Devriendt}, J.~E.~G., {Gibson}, B.~K., \& {Silk}, J. 2003,  \mnras, 345, 1285

\bibitem[{Kogut}, {Spergel}, {Barnes}, {Bennett},  {Halpern}, {Hinshaw}, {Jarosik}, {Limon}, {Meyer}, {Page}, {Tucker},  {Wollack}, \& {Wright} 2003]{kogut03}
{Kogut}, A., {Spergel}, D.~N., {Barnes}, C., {Bennett}, C.~L., {Halpern}, M.,  {Hinshaw}, G., {Jarosik}, N., {Limon}, M., {et al.} 2003, \apjs, 148, 161

\bibitem[{Kormendy} \& {Fisher} 2005]{2005RMxAC..23..101K}
{Kormendy}, J. \& {Fisher}, D.~B. 2005, in Revista Mexicana de Astronomia y  Astrofisica Conference Series, 101--108

\bibitem[{Kravtsov}, {Gnedin}, \&  {Klypin} 2004]{2004ApJ...609..482K}
{Kravtsov}, A.~V., {Gnedin}, O.~Y., \& {Klypin}, A.~A. 2004, \apj, 609, 482

\bibitem[{Machacek}, {Bryan}, \&  {Abel} 2001]{2001ApJ...548..509M}
{Machacek}, M.~E., {Bryan}, G.~L., \& {Abel}, T. 2001, \apj, 548, 509

\bibitem[{Metcalfe}, {Shanks}, {Campos}, {McCracken},  \& {Fong} 2001]{2001MNRAS.323..795M}
{Metcalfe}, N., {Shanks}, T., {Campos}, A., {McCracken}, H.~J., \& {Fong}, R.  2001, \mnras, 323, 795

\bibitem[{Minchin}, {Davies}, {Disney}, {Boyce},  {Garcia}, {Jordan}, {Kilborn}, {Lang}, {Roberts}, {Sabatini}, \& {van  Driel} 2005]{2005ApJ...622L..21M}
{Minchin}, R., {Davies}, J., {Disney}, M., {Boyce}, P., {Garcia}, D., {Jordan},  C., {Kilborn}, V., {Lang}, R., {et al.} 2005, \apjl, 622, L21

\bibitem[{Moore}, {Ghigna}, {Governato},  {Lake}, {Quinn}, {Stadel}, \& {Tozzi} 1999a]{1999ApJ...524L..19M}
{Moore}, B., {Ghigna}, S., {Governato}, F., {Lake}, G., {Quinn}, T., {Stadel},  J., \& {Tozzi}, P. 1999a, \apjl, 524, L19

\bibitem[{Moore}, {Quinn}, {Governato},  {Stadel}, \& {Lake} 1999b]{1999MNRAS.310.1147Mb}
{Moore}, B., {Quinn}, T., {Governato}, F., {Stadel}, J., \& {Lake}, G.  1999b, \mnras, 310, 1147

\bibitem[{Narayanan}, {Spergel}, {Dav{\'e}}, \&  {Ma} 2000]{2000ApJ...543L.103N}
{Narayanan}, V.~K., {Spergel}, D.~N., {Dav{\'e}}, R., \& {Ma}, C.-P. 2000,  \apjl, 543, L103

\bibitem[{Navarro}, {Frenk}, \&  {White} 1997]{1997ApJ...490..493N}
{Navarro}, J.~F., {Frenk}, C.~S., \& {White}, S.~D.~M. 1997, \apj, 490, 493

\bibitem[{Navarro}, {Hayashi}, {Power}, {Jenkins},  {Frenk}, {White}, {Springel}, {Stadel}, \& {Quinn} 2004]{2004MNRAS.349.1039N}
{Navarro}, J.~F., {Hayashi}, E., {Power}, C., {Jenkins}, A.~R., {Frenk}, C.~S.,  {White}, S.~D.~M., {Springel}, V., {Stadel}, J., {et al.} 2004,  \mnras, 349, 1039

\bibitem[{Norman} \& {Bryan} 1999]{norman99}
{Norman}, M. \& {Bryan}, G. 1999, Numerical Astrophysics : Proceedings of the  International Conference on Numerical Astrophysics 1998 (NAP98), held at the  National Olympic Memorial Youth Center, Tokyo, Japan, March 10-13, 1998., ed.  K.~T. S.~M.~Miyama \& T.~Hanawa (Kluwer Academic)

\bibitem[{Omukai} \& {Palla} 2003]{2003ApJ...589..677O}
{Omukai}, K. \& {Palla}, F. 2003, \apj, 589, 677

\bibitem[{O'Shea}, {Bryan}, {Bordner}, {Norman},  {Abel}, \& {Harkness} 2004]{oshea04}
{O'Shea}, B., {Bryan}, G., {Bordner}, J., {Norman}, M., {Abel}, T., \&  {Harkness}, R. amd~{Kritsuk}, A. 2004, Adaptive Mesh Refinement - Theory and  Applications, ed. T.~Plewa, T.~Linde, \& G.~Weirs (Springer-Verlag)

\bibitem[{O'Shea}, {Nagamine}, {Springel}, {Hernquist},  \& {Norman} 2005]{2005ApJS..160....1O}
{O'Shea}, B.~W., {Nagamine}, K., {Springel}, V., {Hernquist}, L., \& {Norman},  M.~L. 2005, \apjs, 160, 1

\bibitem[{Page}, {Hinshaw}, {Komatsu}, {Nolta},  {Spergel}, {Bennett}, {Barnes}, {Bean}, {Dore'}, {Halpern}, {Hill},  {Jarosik}, {Kogut}, {Limon}, {Meyer}, {Odegard}, {Peiris}, {Tucker}, {Verde},  {Weiland}, {Wollack}, \& {Wright} 2006]{2006astro.ph..3450P}
{Page}, L., {Hinshaw}, G., {Komatsu}, E., {Nolta}, M.~R., {Spergel}, D.~N.,  {Bennett}, C.~L., {Barnes}, C., {Bean}, R., {et al.} 2006, ArXiv Astrophysics e-prints

\bibitem[{Peebles} 1968]{1968ApJ...153....1P}
{Peebles}, P.~J.~E. 1968, \apj, 153, 1

\bibitem[{Peebles} 2001]{2001ApJ...557..495P}
---. 2001, \apj, 557, 495

\bibitem[{Percival}, {Baugh}, {Bland-Hawthorn},  {Bridges}, {Cannon}, {Cole}, {Colless}, {Collins}, {Couch}, {Dalton}, {De  Propris}, {Driver}, {Efstathiou}, {Ellis}, {Frenk}, {Glazebrook}, {Jackson},  {Lahav}, {Lewis}, {Lumsden}, {Maddox}, {Moody}, {Norberg}, {Peacock},  {Peterson}, {Sutherland}, \& {Taylor} 2001]{2001MNRAS.327.1297P}
{Percival}, W.~J., {Baugh}, C.~M., {Bland-Hawthorn}, J., {Bridges}, T.,  {Cannon}, R., {Cole}, S., {Colless}, M., {Collins}, C., {et al.} 2001, \mnras,  327, 1297

\bibitem[{Ripamonti} \& {Abel} 2004]{ripamonti04}
{Ripamonti}, E. \& {Abel}, T. 2004, \mnras, 348, 1019

\bibitem[{Schaerer} 2002]{2002A&A...382...28S}
{Schaerer}, D. 2002, \aap, 382, 28

\bibitem[{Sokasian}, {Abel}, {Hernquist}, \&  {Springel} 2003]{2003MNRAS.344..607S}
{Sokasian}, A., {Abel}, T., {Hernquist}, L., \& {Springel}, V. 2003, \mnras,  344, 607

\bibitem[{Somerville}, {Bullock}, \&  {Livio} 2003]{2003ApJ...593..616S}
{Somerville}, R.~S., {Bullock}, J.~S., \& {Livio}, M. 2003, \apj, 593, 616

\bibitem[{Sommer-Larsen} \& {Dolgov} 2001]{2001ApJ...551..608S}
{Sommer-Larsen}, J. \& {Dolgov}, A. 2001, \apj, 551, 608

\bibitem[{Stone} \& {Norman} 1992a]{stone92a}
{Stone}, J.~M. \& {Norman}, M.~L. 1992a, \apjs, 80, 753

\bibitem[{Stone} \& {Norman} 1992b]{stone92b}
---. 1992b, \apjs, 80, 791

\bibitem[{Swaters}, {Madore}, \&  {Trewhella} 2000]{2000ApJ...531L.107S}
{Swaters}, R.~A., {Madore}, B.~F., \& {Trewhella}, M. 2000, \apjl, 531, L107

\bibitem[{Tegmark}, {Blanton}, {Strauss}, {Hoyle},  {Schlegel}, {Scoccimarro}, {Vogeley}, {Weinberg}, {Zehavi}, {Berlind},  {Budavari}, {Connolly}, {Eisenstein}, {Finkbeiner}, {Frieman}, {Gunn},  {Hamilton}, {Hui}, {Jain}, {Johnston}, {Kent}, {Lin}, {Nakajima}, {Nichol},  {Ostriker}, {Pope}, {Scranton}, {Seljak}, {Sheth}, {Stebbins}, {Szalay},  {Szapudi}, {Verde}, {Xu}, {Annis}, {Bahcall}, {Brinkmann}, {Burles},  {Castander}, {Csabai}, {Loveday}, {Doi}, {Fukugita}, {Gott}, {Hennessy},  {Hogg}, {Ivezi{\'c}}, {Knapp}, {Lamb}, {Lee}, {Lupton}, {McKay}, {Kunszt},  {Munn}, {O'Connell}, {Peoples}, {Pier}, {Richmond}, {Rockosi}, {Schneider},  {Stoughton}, {Tucker}, {Vanden Berk}, {Yanny}, \&  {York} 2004]{2004ApJ...606..702T}
{Tegmark}, M., {Blanton}, M.~R., {Strauss}, M.~A., {Hoyle}, F., {Schlegel}, D.,  {Scoccimarro}, R., {Vogeley}, M.~S., {Weinberg}, D.~H., {et al.} 2004, \apj, 606,  702

\bibitem[{Tegmark}, {Silk}, {Rees}, {Blanchard},  {Abel}, \& {Palla} 1997]{1997ApJ...474....1T}
{Tegmark}, M., {Silk}, J., {Rees}, M.~J., {Blanchard}, A., {Abel}, T., \&  {Palla}, F. 1997, \apj, 474, 1

\bibitem[{Truelove}, {Klein}, {McKee}, {Holliman},  {Howell}, \& {Greenough} 1997]{truelove97}
{Truelove}, J.~K., {Klein}, R.~I., {McKee}, C.~F., {Holliman}, J.~H., {Howell},  L.~H., \& {Greenough}, J.~A. 1997, \apjl, 489, L179+

\bibitem[{van den Bosch}, {Robertson},  {Dalcanton}, \& {de Blok} 2000]{2000AJ....119.1579V}
{van den Bosch}, F.~C., {Robertson}, B.~E., {Dalcanton}, J.~J., \& {de Blok},  W.~J.~G. 2000, \aj, 119, 1579

\bibitem[{Viel}, {Lesgourgues}, {Haehnelt}, {Matarrese},  \& {Riotto} 2005]{2005PhRvD..71f3534V}
{Viel}, M., {Lesgourgues}, J., {Haehnelt}, M.~G., {Matarrese}, S., \& {Riotto},  A. 2005, \prd, 71, 063534

\bibitem[{White} 1996]{1996clss.conf..349W}
{White}, S.~D.~M. 1996, in Cosmology and Large Scale Structure, 349--+

\bibitem[{Willman}, {Governato}, {Dalcanton}, {Reed},  \& {Quinn} 2004]{2004MNRAS.353..639W}
{Willman}, B., {Governato}, F., {Dalcanton}, J.~J., {Reed}, D., \& {Quinn}, T.  2004, \mnras, 353, 639

\bibitem[{Woodward} \& {Colella} 1984]{Woodward84}
{Woodward}, P.~R. \& {Colella}, P. 1984, J. Comp. Phys., 54, 174

\bibitem[{Yoshida}, {Abel}, {Hernquist},  \& {Sugiyama} 2003a]{2003ApJ...592..645Y}
{Yoshida}, N., {Abel}, T., {Hernquist}, L., \& {Sugiyama}, N.  2003a, \apj, 592, 645

\bibitem[{Yoshida}, {Sokasian},  {Hernquist}, \& {Springel} 2003b]{2003ApJ...591L...1Y}
{Yoshida}, N., {Sokasian}, A., {Hernquist}, L., \& {Springel}, V.  2003b, \apjl, 591, L1

\end{thebibliography}
\end{document}